\documentclass[acmsmall,screen]{acmart}

\AtBeginDocument{%
  \providecommand\BibTeX{{%
    \normalfont B\kern-0.5em{\scshape i\kern-0.25em b}\kern-0.8em\TeX}}}

\usepackage{makecell}
\usepackage{tabularx}
    \newcolumntype{L}{>{\raggedright\arraybackslash}X}

\renewcommand\footnotetextcopyrightpermission[1]{}
\copyrightyear{}
\acmYear{}
\acmDOI{}
\acmConference[]{}{}{}
\acmBooktitle{}
\acmPrice{}
\acmISBN{}

\begin{document}

\title{Assessing the Impact of Music Recommendation Diversity on Listeners: A Longitudinal Study}

\author{Lorenzo Porcaro}
\email{lorenzo.porcaro@upf.edu}
\affiliation{%
  \institution{Music Technology Group, Universitat Pompeu Fabra}
  \city{Barcelona}
  \country{Spain}
}
\affiliation{%
  \institution{Joint Research Centre, European Commission}
  \city{Ispra}
  \country{Italy}
}

\author{Emilia G\'{o}mez}
\email{emilia.gomez@upf.edu}
\affiliation{%
  \institution{Music Technology Group, Universitat Pompeu Fabra}
  \city{Barcelona}
  \country{Spain}
}
\affiliation{%
  \institution{Joint Research Centre, European Commission}
  \city{Sevilla}
  \country{Spain}
}

\author{Carlos Castillo}
\email{carlos.castillo@upf.edu}
\affiliation{%
  \institution{Web Science and Social Computing Group, Universitat Pompeu Fabra}
  \city{Barcelona}
  \country{Spain}
}
\affiliation{%
  \institution{ICREA}
  \city{Barcelona}
  \country{Spain}
}

\renewcommand{\shortauthors}{Lorenzo Porcaro, Emilia G\'{o}mez, \& Carlos Castillo}

\begin{abstract}
We present the results of a 12-week longitudinal user study wherein the participants, 110 subjects from Southern Europe, received on a daily basis Electronic Music (EM) diversified recommendations.
By analyzing their explicit and implicit feedback, we show that exposure to specific levels of music recommendation diversity may be responsible for long-term impacts on listeners' attitudes. 
In particular, we highlight the function of diversity in increasing the openness in listening to EM, a music genre not particularly known or liked by the participants previous to their participation in the study. 
Moreover, we demonstrate that recommendations may help listeners in removing positive and negative attachments towards EM, deconstructing pre-existing implicit associations but also stereotypes associated with this music.
In addition, our results show the significant clout that recommendation diversity has in generating curiosity in listeners.
\end{abstract}



\received{---}
\received[revised]{---}
\received[accepted]{---}

\maketitle

\section{Introduction}\label{01_intro}
Recommender Systems (RS) affect several choices in our daily life, helping us choose, for instance, the news we read, the movies we watch, the job positions we apply for, or the music we listen to.
Besides the end-users consuming the recommendations, RS also affect those producing the items being recommended:
artists' royalty revenues may depend on whether they are recommended or not within a streaming platform; 
the fame of a brand in e-commerce may rely upon if its products are displayed or not as similar to the ones previously purchased by users; 
the time spent unemployed may depend on whether a profile is shown or not to potential employers.
These are just a few examples of the several RS stakeholders subject to different, sometimes unintended, and emerging impacts \citep{Jannach2019}.

The awareness of these impacts is at the basis of the flourishing fair ranking and recommendations literature \citep{Patro2022}.
For instance, Hasan et al. \cite{Hasan2018} show that RS potentially increases excessive video usage on online platforms. 
Adomavicius et al. \cite{Adomavicius2013} highlight the role of recommendations in manipulating consumers' preference ratings. 
Fabbri et al. \cite{Fabbri2022} investigate the role of RS in promoting users' radicalization. 
Notwithstanding, RS may certainly be responsible also for positive changes. 
Hauptmann et al. \cite{Hauptmann2021} propose an app for healthy food recommendations that positively affect nutritional behaviour. 
Music recommendations have been used for helping recover the musical memory of people with Alzheimer's disease \citep{Navarro2014}. 
In the work by Starke et al. \cite{Starke2021} users' adoption of energy-saving measures is boosted thanks to a proper recommendation interface design. 
Other examples are provided in the comprehensive overview of RS stakeholders, values, and risks by Jannach and Bauer \cite{Jannach2020}.

Among the recommendation characteristics under the spotlight in RS impact-oriented research, diversity has drawn the interest of researchers, practitioners, policy-makers and also affected communities because of its latent influence on individuals' choices. 
In particular, \textit{exposure diversity} as mediated by recommender systems is a research topic attracting scholars from different disciplines, especially in relation to its impact on human rights such as inclusion, non-discrimination and fairness \cite{Helberger2018}.
Previous works in the RS literature have investigated how recommendations may influence aspects such as consumption or sales diversity, sometimes focusing on the impact of diversity on other recommendation characteristics, such as their usefulness or attractiveness e.g., \cite{Willemsen2016}. 
Nevertheless, in this strain of RS research longitudinal user studies are still quite rare.

We contribute to this latter corpus of research with this work, the first longitudinal user study in the Music RS literature presenting an impact analysis of diversity on listeners' attitudes.
It consists of a 12-week study wherein the participants, 110 subjects from Southern Europe, received daily Electronic Music (EM) recommendations with different levels of diversity. 
We propose a critical evaluation that takes into account insights from the music psychology and education field. 
Indeed, music exposure is proven to have the power of reducing stereotypes and prejudice against unknown or unfamiliar cultures. 
In addition, repeated exposure and familiarity have been linked to the construction of aesthetic preferences in the music domain. 
Under this lens, we evaluate the impact of recommendation diversity not from a behavioural perspective but instead assessing how music recommendations may be a vehicle for attitudinal change. 
Specifically, the purpose of this study is to answer the following research questions: 
\begin{itemize}
    \item[\textbf{[RQ1]}]\textbf{to what extent listeners' implicit and explicit attitudes towards an unfamiliar music genre can be affected by exposure to music recommendations?}
    \item[\textbf{[RQ2]}] \textbf{what is the relationship between music recommendation diversity and the impact on listeners' attitudes?}
\end{itemize}

By analyzing participants' explicit and implicit feedback, we show that exposure to specific levels of music recommendation diversity may be responsible for long-term impacts on listeners' attitudes. 
In particular, we highlight the function of diversity in increasing the openness in listening to EM, a music genre not particularly known or liked by the participants previous to their participation in the study. 
Moreover, we demonstrate that recommendations may help listeners in removing positive and negative attachments towards EM, deconstructing pre-existing implicit associations but also stereotypes associated with this music.
In addition, our results show the significant clout that recommendation diversity has in generating curiosity in listeners.

The rest of the article is organised as follows. 
Section \ref{02_background} starts with an overview of impact assessment practices, followed by a brief survey of recent developments of RS simulation-based methods, and lastly presents several works on the impact of music exposure on listeners. 
Afterwards, Section \ref{03_studydesign} describes the user study design, while Section \ref{04_material} reports the process designed to create the music recommendation to which study participants have been exposed. 
Then, Section \ref{05_results} presents the results of our analysis, discussed together with their limitations in Section \ref{06_discussion}. 
Finally, conclusions are drawn in Section \ref{07_conclusions}.

\section{Background and Related Work}\label{02_background}
Algorithmic Impact Assessment (AIA) is a complex process that goes beyond the development of practices to measure quantitatively some kind of change. 
Instead, it includes the involvement of several actors, starting from the system designers arriving at the community affected by algorithmic systems. 
In Section \ref{02_AIA}, we discuss the idea of \textit{impact} and \textit{impact assessment}, in particular with regard to recent proposals of AIA. 
Afterwards, in Section \ref{02_simulation} we centre our attention on RS simulation-based frameworks, which have attracted the attention of the practitioners interested in assessing the impact of these systems. 
On the contrary, as recently observed by Liang and Willemsen \cite{Liang2022b}, longitudinal studies are not common in RS literature and their recent work is a notable exception, together with the study presented by Hauptmann et al. \cite{Hauptmann2021}.
Finally, in Section \ref{02_music_exposure} we present several insights from the music psychology field on the impact of music exposure on listeners.

\subsection{Algorithmic Impact Assessment} \label{02_AIA}
In its simplest form, Impact Assessment (IA) may be defined as ``[...] the process of identifying the future consequences of a current or proposed action. The \textit{impact} is the difference between what would happen with the action and what would happen without it'' \cite{IAIA2009}.
Far from being a novel area of research, IA practices have been discussed globally since 1980 by the International Association for Impact Assessment (IAIA),\footnote{\url{https://iaia.org}} and justifiably one among the first framework proposed in this regard is the \textit{Environmental Impact Assessment} (EIA). 
During the last 40 years, EIA had to deal with issues associated with loss of biodiversity, damage to marine areas, and climate change, just to mention a few \citep{Morgan2012}.

At the same time the second strain of IA practices has emerged to monitor socio-cultural impacts, among the others \textit{Social} (and \textit{Societal}) \textit{Impact Assessment} (SIA) \citep{Vanclay2003, Vanclay2011, Kreissl2015}, \textit{Human Rights Impact Assessment} (HRIA) \citep{Kemp2013}, and \textit{Cultural Impact Assessment} (CIA) \citep{Partal2016}.
Vanclay \cite{Vanclay2003} defines SIA as ``analysing, monitoring and managing the social consequences of development'', wherein a particular focus is posed on values such as fairness and equity, commitment to sustainability, or openness and accountability \citep{Vanclay2011}. 
Similarly, Societal Impact Assessment targets technology as potentially responsible for altering society as a whole, and Kreissl et al. \cite{Kreissl2015} exemplify this framework by discussing the case of Privacy Impact Assessment (PIA).
Instead, HRIA is based on a series of internationally recognised human rights, such as the right to a livelihood, the right to participate in the cultural life of a community, or the right to a fair wage \citep{Kemp2013}.

Algorithmic Impact Assessment (AIA) is relatively a new field in comparison to the aforementioned frameworks, but nevertheless, its importance is quickly growing. 
Despite its relations with other IA such as Privacy Impact Assessment (PIA) or Data Protection Impact Assessment (DPIA)  \citep{Kaminski2021}, AIA can be defined as a set of ``emerging governance practices for delineating accountability, rendering visible the harms caused by algorithmic systems, and ensuring practical steps are taken to ameliorate those harms'' \citep{Metcalf2021}. 
A few aspects of AIA are addressed in detail below, but we point the reader interested in a comprehensive overview and discussion on current AIA practices towards the reports published by the non-profit organisations \textit{AI Now Institute} \citep{Reisman2018} and \textit{Data \& Society} \citep{Moss2021}.

As discussed by Vecchione et al. \cite{Vecchione2021} in the context of algorithmic auditing, most of the impact which may result from the interaction with an algorithmic system appears beyond discrete moments of decision-making. 
This is particularly true for RS, wherein the impact may not be evident after a single interaction, but instead be the fruit of multiple exposures through time. 
Second, as Metcalf et al. \cite{Metcalf2021} argue, the impact is co-constructed by all the actors linked to an algorithmic system: developers, designers, decision-makers, public and private organisations, and most importantly the affected communities. 
Therefore, it is fundamental to involve each of these actors while defining the AIA practice, a vision shared also in \cite{Vecchione2021}. 
While the exploration of the long-term impact is already on the agenda of RS practitioners, and as discussed in the next section mostly addressed by the use of simulated environments, the involvement of the wider community of people affected by RS is still rare to find. 
A notable exception is a work by Ferraro \cite{Ferraro2021} considering the artists' perspective on the impact of music recommendation, wherein he interviews several artists to understand how effectively algorithmic automated decisions were impacting their work-life. 
In the music field, another example that is worth mentioning is the work done in the Algorithmic Responsibility research area at Spotify, resulting in the development of AIA in the platform \citep{Spotify2022}.

A limitation of current impact-oriented RS research is the lack of assembling a wide range of expertise. 
Indeed, being RS research mostly driven by computer science-inspired approaches, most of what until now has been measured as impact is the result of technical and engineering knowledge.
Despite the several efforts to properly develop robust metrics and evaluation procedures, the narrowness of the considered approaches has been counterproductive because it limited the concept of impact itself to what is measurable according to such procedures, a deterministic approach by definition which needs to be expanded to make AIA practices shared and effective.

\subsection{Diversity in Simulation-based Methods}\label{02_simulation}
The interest in impact-oriented research using synthetic data and simulation environments is rapidly growing within and outside the RS community \citep{Ekstrand2021b}. 
However, this growing interest is accompanied by an equally growing concern about the high heterogeneity and low transparency of methods, evaluation practices, and more general assumptions on which such simulation studies are built \citep{Winecoff2021}. 
Next, we review the simulation-based recommender system literature wherein the \textit{impact on}, or the \textit{impact of} diversity has been studied. 

Agent-Based Modelling (ABM) has been applied in several works to understand the long-term impact of RS \cite{Adomavicius2021}.
For instance, Zhang et al. \cite{Zhang2020} centre their attention on simulating users' consumption strategies, showing how relying on recommendations users may end in the long-term contributing to the decrease of aggregate diversity. 
Zhou et al. \cite{Zhou2021} use ABM to study how \textit{preference bias} --- the distortion in users' self-reported ratings caused by recommendations --- influences the performance of recommender systems.
In particular, they show how the bias introduced into the system through the users' ratings may negatively influence the overall diversity of the recommended items.

Further examples of simulated environments can be found in the RS literature analysing the impact of feedback loops.
Mansoury et al. \cite{Mansoury2020} design an iterative model to analyse the feedback loop, showing how it may cause a decline in aggregate diversity.
Moreover, Jiang et al. \cite{Jiang2019} provide a theoretical analysis of the relationship between feedback loops, echo chambers and filter bubbles. 
Instead, Chaney et al. \cite{Chaney2018} by simulating different models of users' engagement with RS prove the impact of feedback loops on algorithm confounding, indicating such outcome as the cause of homogenization of users' behaviours.

Another strain of simulation-based analysis focuses on comparing the performance of different typologies of recommender systems. 
Hazrati et al. \cite{Hazrati2020} create a simulated environment where users are exposed to different kinds of recommender systems, proving that non-personalized methods produce the lowest diversity in terms of users' choices. 
Ferraro et al. \cite{Ferraro2020} focus on the analysis of several session-based recommendation techniques in the music domain, using an iterative approach wherein users are assumed to interact with a fraction of the recommended tracks.
The authors show that in terms of the spread and coverage of recommendations, the various systems analysed led to an increased concentration effect over time. 
The same simulation technique is presented by Jannach et al. \cite{Jannach2015}, this time in the context of movie recommendations, where again the authors show how different algorithms lead to highly different performances in both recommendations spread and coverage.

Under a different lens, \textit{sales diversity} is at the centre of attention in a series of studies by Fleder and Hosangar \cite{Fleder2007, Fleder2009} and Lee and Hosanagar \cite{Lee2014, Lee2017}. 
Fleder and Hosangar prove that collaborative filtering recommender systems exert a \textit{concentration bias}, early delineating the idea that it may be possible that, at the individual level, diversity may increase, while at the aggregate level, diversity may decrease. 
Lee and Hosanger using field experiment data confirm such results, highlighting how users do effectively explore novel items thanks to the recommendations, but such explorations are highly correlated among users.

The impact on \textit{content} and \textit{source} \textit{diversity} has been also investigated.
Haim et al. \cite{Haim2018} model four agents interacting with a personalised news recommender system in order to study the effect of personalization on diversity. 
They do no evidence of any strong link between personalization and content and source diversity, as similarly found in \cite{Moller2018}.
Moreover, Aridor et al. \cite{Aridor2020} use numerical simulations to model user decision-making processes, providing an explanation of the findings of a previous study by Nguyen et al. \cite{Nguyen2014} on the influence of recommender systems on content diversity.
In the latter, the authors found that users' interacting with the provided recommendations ended up consuming more diverse content in comparison to the users who did not. 
Aridor and colleagues confirm such results by means of their simulation, but also observe an increase in the homogeneity across users i.e. decrease in aggregate diversity.

\subsection{Impact of Music Exposure}\label{02_music_exposure}
Several scholars in the music psychology and education field investigated the role of music exposure in influencing stereotypes and attitudes, and we highlight some examples hereafter.

Greitemeyer and colleagues \cite{Greitemeyer2014} prove that exposure to music with pro-integration lyrics may reduce prejudice and discrimination towards immigrant groups, and later in \citep{Greitemeyer2015} they show that exposing listeners to music with pro-equality lyrics may enhance positive attitudes and behaviours toward women. 
In both cases, the authors found that the musical characteristics of the music exposed and the preference for it do not influence such impacts. 
Clarke et al. \cite{Clarke2015} investigate the relationship between music, empathy and cultural understanding, showing that music exposure may indeed generate a sense of affiliation with unknown cultures. 
Their results are expanded in \citep{Vuoskoski2017}, confirming that, especially in listeners with high trait empathy, music may increase positive implicit attitudes towards images representing members of foreign cultures. 
Tu \cite{Tu2009} provides empirical evidence that exposing young students to 10 minutes of a Chinese music curriculum, when prolonged for 10 weeks, may impact the attitudes towards Chinese people. 
In a similar study, Sousa et al. \cite{Sousa2005} show that exposure to Cape Verdean songs together with Portuguese songs may reduce anti-dark-skinned stereotyping among light-skinned Portuguese children. 
Even if these examples consider different stimuli and different subjects, they all provide evidence that music listening may impact the idea that we have about other social groups, and more broadly about other cultures.

Another strain of research in the field of Music Psychology has been interested in understanding the impact of repeated exposure on music preference, based on Berlyne's psychobiological theory (see \cite{Chmiel2017} for an overview). 
Among his several contributions, he theorised the existence of a relationship between aesthetic preferences and familiarity. 
Even if some studies support his claims, e.g., \citep{Szpunar2004}, while others confute it, e.g., \citep{Madison2017}, nowadays it is commonly accepted that familiarity with music has a prominent role in the development of preferences \citep{Johnston2021}, a topic in which also neuroscience practitioners extensively debate \citep{Freitas2018}. 
These studies motivate our interest in exploring the impact that music recommendation diversity may have, connecting exposure, familiarity, and preferences. 
Indeed, whilst music recommender systems are known to be influential on exposure diversity \citep{Helberger2018, Porcaro2021}, less is known about how such exposure diversity may affect the users' opinions, beliefs and attitudes.

\section{Study Design}\label{03_studydesign}
The study is divided into four main stages, namely prescreening, \textit{PRE}, \textit{COND} and \textit{POST}, detailed hereafter.
First, we designed a prescreening online survey to select subjects matching a set of criteria, presented in Section \ref{03_prescreening}.
After selecting the desired set of participants, we started collecting their data using two methods. 
First, we asked them to create a ListenBrainz\footnote{\url{https://listenbrainz.org}} account to gather information about their listening habits (Section \ref{03_listenbrainz}). 
Additionally, participants completed the Electronic Music Feedback (EMF) questionnaire, where they provided their opinion about several aspects of EM (Section \ref{03_EMF}). 
For the following four weeks, no further actions were required to the participants. This stage of the experiment is referred to as \textit{PRE}.

\begin{figure}[h!]
\centering
\includegraphics[width=0.9\textwidth]{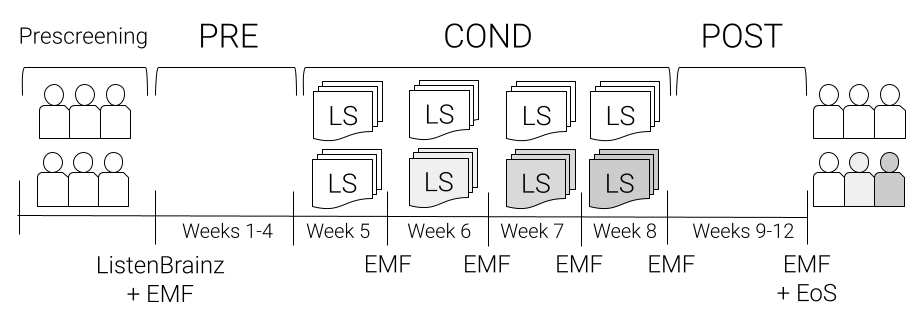}
\caption{High-level view of the longitudinal study. Participants are divided into two groups: low diversity recommendations (\textit{white}) and high diversity recommendations (\textit{shading grey}). LS stands for Listening Sessions. EMF stands for Electronic Music Feedback questionnaire. EoS stands for End-of-Study survey.}
\label{fig:5_1}
\end{figure}

In the 5th week, participants started to be exposed to music recommendations in what we call the \textit{COND} (conditioning) stage. 
At that point, participants were already randomly divided into two groups, one receiving high diversity (HD) and the other low diversity (LD) recommendations, created following the procedure described in Section \ref{04_material}.
For four weeks, from Monday to Friday, participants received on a daily basis an audio mix to be listened to, for a total of 20 listening sessions (Section \ref{sec:03_listsessions}). 
After each listening session, additional feedback was collected by asking participants their impressions about the music listened to. 
During this stage, they were also asked to complete the EMF questionnaire on a weekly basis, on Saturday. 
At the start of the 9th week, the \textit{COND} stage ended and the \textit{POST} stage started. 
Again for four weeks, no further actions were required to the participants.
Finally, at the end of the 12th week, we asked participants first to fill for the last time the EMF questionnaire, and then to fill out the End-of-Study (EoS) survey described in Section \ref{sec:03_final}.
Figure \ref{fig:5_1} depicts the study's high-level structure. 

\subsection{Recruitment and Informed Consent}
The Institutional Committee for Ethical Review of Projects (CIREP) at Universitat Pompeu Fabra approved the study design and confirmed the compliance of the research project with the data protection legal framework, namely, with the European General Data Protection Regulation (EU) 2016/679 (GDPR) and Spanish Organic Law 3/2018, of December 5th, on Protection of Personal Data and Guarantee of Digital Rights (LOPDGDD).
A digital copy of the submitted documentation, the ethics certificate and the data protection certificate is available upon request.

Before the main study, a smaller-scale pilot study was conducted to test the data collection process. 
The pilot study took place in February and April 2022, and the main study took place from May to July 2022. 
All participants were recruited using the online recruitment service \textit{Prolific},\footnote{\url{https://prolific.co}} and they were paid \pounds 6.00 per hour, the recommended minimum pay. 
They were informed about the voluntary nature of their participation, having the freedom to withdraw at any point, and of their rights including the right to access, rectify, and delete their information. 
They were also shown the information sheet describing the research objectives, methodology, risks, and benefits. 
Informed consent was obtained from all participants.


\subsection{Prescreening}\label{03_prescreening}
The prescreening of the participants was performed as a two-step process: first, using pre-determined criteria that are available in the recruiting platform (Prolific), and then, based on a questionnaire.

In the first step, we selected participants based on age (18-42), nationality and country of residence (Italy, Spain, and Portugal), fluent in English, who had participated in at least 20 surveys in Prolific, and with a task approval rate above 90\%.
In terms of gender, sex, and education level, no filter was applied. 
We chose to limit age including only Millennials and Generation Z subjects, i.e., those born between 1981 and 2012, known to have a predilection for Electronic Music (EM) \citep{Nielsen2014, Voigt2016, Luminate2022}, but also to narrow the generational differences among participants. 
The selection of only three countries in Southern Europe was motivated by the idea of having participants: 
1) with a relatively similar cultural background; 
2) living in the same time zone (GMT +1/+2). 
This last factor was fundamental to facilitating the daily interaction between participants and the researchers, being they resident in the same time zone.

Subjects matching the aforementioned criteria were redirected to the second part of the prescreening that consisted of a questionnaire in \textit{PsyToolkit} \citep{Stoet2010, Stoet2017}, a web-based framework to conduct psychological surveys and experiments.
The questionnaire was composed of three main parts. 
In the first part, we asked participants to optionally confirm their demographic information. 
This step was included to double-check the reliability of the information provided by Prolific.
In the second part, we asked for additional information about participants' listening habits. 
In detail, they self-assessed with 5-point Likert items their taste variety, EM listening frequency, and EM taste variety. 
Additionally, we asked them to indicate the preferred music streaming platform and the average daily time spent listening to music. 
This information was used to filter out participants who self-declared to listen to EM very often, who indicated listening to music less than an hour a day, and who did not select Spotify as the preferred streaming service. 
The latter condition was necessary for collecting participants' listening logs through ListenBrainz, as explained in the next section. 
The former two conditions were designed to create a group of subjects who listen to music more than occasionally, but who are not EM frequent listeners.

In the third part, we included a test to verify the participants' familiarity with EM artists and genres. 
We replicated the test proposed in a previous study \cite{Porcaro2022}, summarised hereafter. 
Participants had to specify whom from a list of mainstream EM artists was (i) known, (ii) possibly known or (iii) unknown to them. 
The list was composed of 20 artists selected by \textit{AllMusic}, an expert-curated online music database, as representatives of EM \cite{AllMusic2022}.
Afterwards, they had to do the same task for a list of EM genres, composed of 20 genres part of the Wikipedia page about EM \cite{Wikipedia2022}.
The final score of each test, separately for artists and genres, was computed giving more points to participants who knew less popular artists (or genres). 
The rationale behind this is that knowing a popular EM artist or genre (such as Skrillex, or \textit{dubstep}) makes you less a connoisseur of EM at large, in comparison to knowing a less popular EM artist or genre (such as Autechre, or \textit{IDM}). 
The popularity of each artist and genre was computed using several signals from \textit{Spotify}, \textit{Twitter}, \textit{Facebook}, \textit{Deezer}, \textit{SoundCloud}, and \textit{Last.fm}, aggregated using the GAP0 metric proposed in \cite{Koutlis2020}. 
The list of artists and genres and the corresponding GAP0 score is presented in Appendix B (Figures \ref{fig:5_2} and \ref{fig:5_3}). 
We filter out participants who according to this test were too familiar with EM. 
In summary, the following criteria were used for the prescreening:
\begin{itemize}
    \item \textbf{First prescreening step}
        \begin{itemize}
            \item \textit{Age}: 18-42
            \item \textit{Nationality \& country of residence}: Italy, Spain, Portugal
            \item \textit{Gender and sex}: No restrictions
            \item \textit{Highest education level}: No restrictions
            \item \textit{Fluent languages}: English
            \item \textit{Number of Prolific previous submissions}: $>$ 20 
            \item \textit{Prolific approval rate}: 90\%
        \end{itemize}
    \item \textbf{Second prescreening step}
        \begin{itemize}
            \item \textit{Taste variety (self-declared)}: No restrictions
            \item \textit{EM taste variety (self-declared)}: No restrictions
            \item \textit{EM listening frequency (self-declared)}: Not very often 
            \item \textit{Preferred music streaming platform}: Spotify
            \item \textit{Average daily listening time (self-declared)}: $>$ 1 hour
            \item \textit{Average EM familiarity score}: $<$ 5 (over 10.5)
        \end{itemize}
\end{itemize}
This prescreening allowed us to select a population of listeners quite homogenous in terms of demographics, listening habits and familiarity with EM. 
Indeed, our main goal is to study the impact of EM recommendations on people who are not experts nor huge fans of this genre. 
We also aimed at reducing the response variability caused by different cultural backgrounds.
These two aspects, familiarity with EM and cultural background, have been shown to be at the root of different perceptions of diversity in music lists \cite{Porcaro2022}, and by controlling for those while selecting the study participants, we aimed at minimizing the influence of such factors in the analysis.

\subsection{Listening Logs Collection}\label{03_listenbrainz}
After being selected for participating in the study, participants were asked to create an account on ListenBrainz, allowing the collection of their listening logs for the entire duration of the study. 
ListenBrainz is a platform that keeps track of what music its users listen to and provides them with insights into their listening habits. 
It is operated by the \textit{MetaBrainz} foundation,\footnote{\url{ https://metabrainz.org}} a non-profit organisation that has been set up to build community-maintained databases and make them available in the public domain or under Creative Commons licences. 
Data is collected complying with the GDPR, and more information about MetaBrainz privacy policy can be found online.

Among the options for submitting the music listened to, it is possible to link Listenbrainz to the Spotify account. 
One of the advantages of this approach is the reliability of the metadata accessible for each log. 
Indeed, once a log is collected by ListenBrainz through its link with Spotify, the associated Spotify track ID and artist ID are available. 
With the retrieved IDs, by using the Spotify Web API,\footnote{\url{https://developer.spotify.com}} it is possible to obtain several types of data, from the acoustic properties of a track to the genres associated with the artists. 
However, a few drawbacks of this method to collect listening logs are noticeable.

First, creating a ListenBrainz account is a time-consuming task for which participants need to be paid, increasing the cost of the experiment. 
Moreover, for privacy reasons people may be reluctant to link their ListenBrainz and Spotify accounts. 
Lastly, whilst the use of the Spotify API is quite accepted in the Music IR and RS communities, the proprietary nature of the algorithms behind the API makes it difficult to know exactly how the data is generated. 
Nonetheless, by using ListenBrainz we aimed to foster the reproducibility of our study, but also to ensure the availability of the collected data for future works making them publicly available through the ListenBrainz API.

\subsection{Electronic Music Feedback Questionnaire}\label{03_EMF}
The Electronic Music Feedback (EMF) questionnaire is designed to measure implicit and explicit attitudes towards EM.
Participants completed it at the beginning, four times while being exposed to the music recommendations, and the last time at the end of the study, for a total of six times (see Figure \ref{fig:5_1}). 
In particular, the EMF measures: 
1) the participants' openness in listening to EM; 
2) the valence of their implicit association with EM; 
3) the stereotypes they associated with EM.
It is implemented in PsyToolkit, and the time needed to complete it is approximately 10 minutes. 
We now continue describing separately the three parts of the questionnaire. 

\subsubsection{Measuring Openness}
Openness in listening to EM is measured using a dichotomous Guttman scale \citep{Guttman1944}. 
It is a unidimensional ordinal cumulative scale for the assessment of an attribute, in this study namely the openness in listening to EM. 
In detail, subjects are asked if they would be open to listening to one hour of EM, selecting \textit{Yes} or \textit{No} to the following options: 
a) once every month; b) once every two weeks; c) once a week; d) twice a week; e) every day. 
The ordinal nature of the scale suggests that a participant answering \textit{No} to the first option, naturally would answer \textit{No} to the following questions, as shown in Table \ref{tab:guttman}.
The score of this scale ranges from 0 for participants declaring to be not open to listening to even one hour per month of EM, to 5 for participants affirming to be open to listening to EM one hour every day. 
Among the advantages of using the Guttman scale are its compact form and pretty intuitive nature while analysing the scores, apart from the ease of computing the score by simply looking for affirmative answers from the participants. 
Having an interest in understanding how music recommendations with different levels of diversity may affect the participants' openness in listening to EM, we used the score obtained from the Guttman scale as one of the variables of the longitudinal analysis, for the rest of the paper referred to as \textbf{o-score}.
\begin{table}[h!]
\caption{Guttman scale built using the question ``Would you be open to listening to one hour of Electronic Music'' for measuring the openness in listening to EM. 0 represent a negative answer, while 1 is an affirmative answer.}
  \centering
  \scalebox{1}{
  \begin{tabular}{ccccc|c}
  \toprule
   \textbf{monthly} & \textbf{biweekly} & \textbf{weekly} & \textbf{twice a week} & \textbf{every day} & \textbf{\textit{o-score}}\\
   \midrule
    0 & 0 & 0 & 0 & 0 & \textbf{0} \\
    1 & 0 & 0 & 0 & 0 & \textbf{1} \\
    1 & 1 & 0 & 0 & 0 & \textbf{2} \\
    1 & 1 & 1 & 0 & 0 & \textbf{3} \\
    1 & 1 & 1 & 1 & 0 & \textbf{4} \\
    1 & 1 & 1 & 1 & 1 & \textbf{5} \\
    \bottomrule
  \end{tabular}}
  \label{tab:guttman}
\end{table}

\subsubsection{Measuring Implicit Association}
People's conscious judgement represents only a facet of how evaluative associations are experienced. 
This is why we included in the questionnaire the Single Category Implicit Association Test (SC-IAT) \citep{Karpinski2006}, a variant of the more famous Implicit Association Test (IAT) \cite{Greenwald1998}. 
The IAT aims at measuring implicit attitudes, defined as ``actions or judgments that are under the control of automatically activated evaluation, without the performer's awareness of that causation'' \citep{Greenwald1998}. 
By measuring the response latencies in a categorization task, the IAT evaluates the strengths of associations between concepts, using complementary pairs of concepts and attributes. 
For instance, IAT has been used to measure people's positive or negative associations with Women and Men, Black and White people, or Transgender and Cisgender people. 
Several examples of IATs can be found on the Project Implicit webpage.\footnote{\url{ https://www.projectimplicit.net}}

In the music field, Clarke, Vuoskoski and DeNora \cite{Clarke2015, Vuoskoski2017} made use of the IAT for measuring if mere exposure to music may evoke empathy towards unknown cultures. 
Their findings support the hypothesis that, even without any accessible semantic content, listening to music can evoke empathy and affiliation in listeners. 
Inspired by their results, we chose to implement an SC-IAT to understand if exposure to EM may influence the implicit association of listeners.
The use of SC-IAT rather than IAT has been motivated by the absence of a complementary category to EM.

In summary, using the keyboard participants have been asked to categorise as fast as possible:
1) pleasant words (\textit{Joy}, \textit{Love}, \textit{Peace}, \textit{Wonderful}, \textit{Pleasure}, \textit{Glorious}, \textit{Laughter}, \textit{Happy}); 
2) unpleasant words (\textit{Agony}, \textit{Terrible}, \textit{Horrible}, \textit{Nasty}, \textit{Evil}, \textit{Awful}, \textit{Failure}, \textit{Hurt}); 
3) EM genres (\textit{Dubstep}, \textit{Techno}, \textit{Electronica}, \textit{Hardcore}, \textit{Vaporwave}, \textit{Breakbeat}, \textit{Electroacoustic}, \textit{Downtempo}). 
By measuring the time they employed in categorising these words correctly, we evaluated participant associations' valence towards Electronic Music. 
The outcome of this test is referred to as \textbf{d-score}, which takes negative values if a negative valence is associated with EM and positive values in the opposite case.\footnote{The test is accessible online: \url{https://github.com/LPorcaro/longterm-musdiv/tree/main/material/SC-IAT}}
Karpinski and Steinman \cite{Karpinski2006} provide a detailed description of the SC-IAT design, the formula for computing the d-score, and proof of its reliability and validity.

\subsubsection{Measuring Stereotypes}
The goal of this part of the EMF questionnaire is to measure what listeners opine on three kinds of stereotypes: 
1) the context wherein they listen to EM; 
2) the musical properties they associate with EM tracks; 
3) the characteristics of EM artists they think are prominent. 
Responses are collected by using 5-point Likert items. 
First, participants are asked in which contexts they would listen to EM presenting a list of eight activities (\textit{Relaxing}, \textit{Commuting}, \textit{Partying}, \textit{Running}, \textit{Shopping}, \textit{Sleeping}, \textit{Studying}, \textit{Working}), selecting an option ranging from \textit{Totally Disagree} to \textit{Totally Agree}. 
In order to analyse the stereotypes associated with EM tracks' musical properties, we ask questions about: 
a) \textit{tempo} (0: mostly slow, 5: mostly fast), 
b) level of \textit{danceability}, 
c) presence of \textit{acoustic instruments}, e.g. violin, trumpet, acoustic guitar , and 
d) presence of \textit{singing voice parts} (0: mostly low, 5: mostly high). 
The reason why we selected these four features is twofold. 
First, they exemplify some of the stereotypes usually associated with EM, e.g., it has a fast tempo, high danceability, and low acousticness. 
Second, these features are among the ones retrievable at a track level (see Appendix \ref{A_recdes}).

Lastly, participants' feedback on which characteristics they associate with Electronic Music artists is collected, focusing on: 
\textit{gender} (0: mostly women or other gender minorities, 5: mostly men); 
\textit{skin colour} (0: mostly white-skinned, 5: mostly dark-skinned); 
\textit{origin} (0: mostly low-income / developing countries, 5: mostly high-income / developed countries); 
and \textit{age} (0: mostly under 40, 5: mostly over 40). 
Considering the nature of the experiment, no impact on the answers related to the artists was expected to be caused by the exposure, because no information about the artists was provided during the listening session.
Nevertheless, understanding the EM artists' characteristics that participants felt to be more representative is a complementary perspective on the stereotypes they associated with EM. 
Even if not influenced by the provided music recommendation, these questions gave us further insights into what ideas the participants had about EM.

\subsection{Listening Sessions}\label{sec:03_listsessions}
Participants' exposure to EM recommendations took place during the twenty daily listening sessions part of the \textit{COND} stage. 
Being the sessions proposed on a daily basis, their design favoured the easiness and rapidness of completing the proposed task, described as follows. 
As an initial step, a thirty-second audio clip was presented to calibrate the audio volume, to avoid exposing participants to extremely louder audio potentially damaging their hearing.
Afterwards, we explicitly asked participants to entirely listen to a 3 minutes audio containing a mix of a few excerpts of EM tracks, asking them to be sure to be in a quiet environment and to allow themselves to be immersed in the music.

\begin{figure}[h!]
\centering
\includegraphics[width=0.8\textwidth]{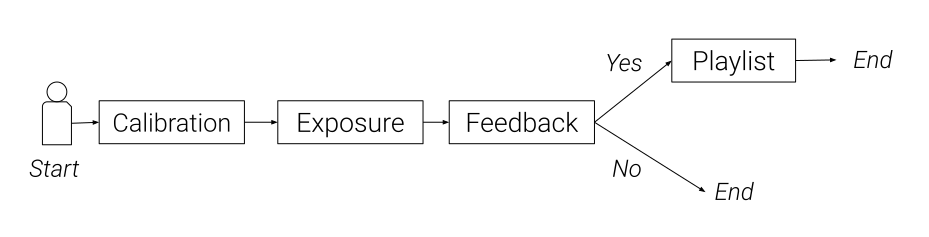}
\caption{Flow diagram representing a listening session.}
\label{fig:5_4}
\end{figure}

After each listening section, the participants provided their feedback on the music listened to.
Specifically, they indicated with a 5-point Likert item if they liked or disliked the music, and they selected if the listened music was familiar or not. 
With that, the mandatory part of the listening session ended. 
Following, on a voluntary basis, they were asked if they wanted to explore the playlist with the full tracks part of the audio previously listened to. 
If selecting \textit{Yes}, they were redirected to a page with a link to a YouTube playlist. 
If they were not interested in discovering, the listening session ended. 
The time needed to complete the mandatory part of each session was approximately 5 minutes.
It is important to note that the interaction with the playlists was declared to be completely optional and did not affect the participants' payment, 
i.e. they were not paid for the time extra spent interacting with the playlists. 
Figure \ref{fig:5_4} depicts the structure of a listening session. 

This design allowed us to collect several types of data. 
First, the explicit liking and the familiarity ratings of the tracks listened to. 
Second, the willingness of discovering more about such tracks, by choosing to explore the playlists. 
Third, by checking the YouTube playlist views, we also had a further metric of the actual interaction with the music proposed daily. 
Whilst the definition of ``legitimate view'' in YouTube is not transparent \citep{Youtube2022}, a common belief is that to increase views a user has to click the play button to begin the video, and the video has to be played for at least 30 seconds. 
Even if we cannot validate these hypotheses, we double-checked that simply accessing a YouTube playlist without listening to any tracks does not increase the view count of such playlists.
We chose to avoid redirecting participants to Spotify playlists to not confound the listening log collections and the platform usage. 
Indeed, Spotify algorithms could have registered the signal of participants' interaction with such playlists influencing eventually future recommendations by including EM related to the study. 
Lastly, YouTube playlists were not made public but accessible only to the participants of the experiment, to avoid affecting the view count with interactions from external users.

\subsection{End-of-Study Survey}\label{sec:03_final}
Over the course of the study, we collected several types of feedback, implicit and explicit, quantitative and qualitative, that we used to understand the impact of music recommendation diversity.
As the last step, we included a final survey to collect participants' overall feedback on their participation. 
It is composed of four sections. 
The first section (16 Likert items) includes questions about the participants’ relationship with EM before participating in the study. 
The second (16 items) collects feedback about the experience of participants during the study. 
In the third (21 items), we ask about participants' feelings about EM after the study. 
Lastly, we insert a final section (8 items) to understand the overall impression of participating in the study.
The complete list of items is presented in the supplementary material.\footnote{\url{https://github.com/LPorcaro/longterm-musdiv/blob/main/material/surveys/5_EndofStudySurvey.pdf}}

The survey is built to investigate the participants' openness, appreciation and willingness to discover EM. 
Besides, a few questions are related to stereotypes associated with this genre (e.g., EM has a fast tempo or it is mostly for partying), while others are about the perceived variety of the genre itself. 
In order to avoid acquiescence bias we balanced the number of positive and negative statements in each section, and items were presented in a randomised fashion to avoid order effect bias. 
The time needed to complete this survey is approximately 10 minutes.

\section{Material}\label{04_material}
This section outlines the semi-automatic procedure to create the diversified recommendations to which participants were exposed during the listening sessions. 
We report here a summary of the several steps carried out, including in Appendix \ref{A_recdes} a detailed description of the process.

The first step was to create a dataset of candidate EM tracks to be included in the listening sessions. 
The goal was to select a set of tracks covering as many EM styles as possible to create a varied representation of the EM culture, however without the presumption of including every existing nuance of it. 
We consulted \textit{Wikipedia} \cite{Wikipedia2022} and \textit{Every Noise at Once}\footnote{\url{https://everynoise.com}} (ENO) to select 20 EM genres with associated 165 subgenres, listed in Table \ref{tab:05_genres}. 
For each of the subgenres, we retrieved from ENO a playlist of around 100 representative tracks for a total of around 16 thousand tracks.
Then, we filtered out tracks too popular by using the Spotify popularity indicator and the YouTube view count, to avoid that familiarity with the music listened to could have an effect on participants' ratings.
Lastly, we randomly select 10 tracks (when available, if not less) for each subgenre, remaining with a final set of 1444 candidate tracks.

As a second step, we made use of Music Information Retrieval (MIR) techniques to design a diversification model.
This was a three-step process: 
1) we extracted audio embedding from the candidate tracks using state-of-the-art Deep Learning models; 
2) we validated these models by using standard MIR hand-crafted features;
3) we designed the diversification process to obtain EM recommendations.

Tracks' audio embeddings were extracted using EfficientNet \cite{Tan2019} trained with a dataset of tracks annotated with Discogs\footnote{\url{https://discogs.com}} metadata.
Among the four models tested, this showed the best performance in terms of clustering the candidate tracks coherently with respect to the considered taxonomy of EM genres. 
Figure \ref{fig:5_6} (Appendix \ref{B_AddTabFig}) displays a 2-dimensional representation of the embedded space obtained with this model.
Then, focusing on four features (\textit{tempo}, \textit{danceability}, \textit{acousticness} and \textit{instrumentalness}) we investigated the consistency of the tracks' embedded space. 
In particular, we centre our attention on if track embeddings clustered together displayed similarity also in terms of the aforementioned features.

We continued by creating two sets of 20 recommendation lists, one list for each listening session, the first set with high and the second with low \textit{inter-list} diversity, from now on the High Diversity (HD) and Low Diversity (LD) lists. 
In the set of HD lists, the idea was to have tracks spanning different EM, giving to the listeners the opportunity to discover different facets of this culture throughout the study. 
In order to do that, we create one list for each of the 20 genres in the dataset. 
On the contrary, LD lists were focused on a single genre (\textit{trance}) for having 20 listening sessions quite homogeneous. 
Figure \ref{fig:5_9} provides a 2-dimensional representation of the recommendation lists obtained using the two diversification strategies (HD and LD).
Besides, every list was formed by four tracks having the minimum average distance in the embedded space, to minimise the \textit{intra-list} diversity. 
Thanks to this, we ensured that each session was coherent and pleasant to listen to, avoiding to include in the same session tracks with very different musical traits.

Once having the 20 HD and 20 LD recommendation lists, each one formed by 4 tracks, the last step was to create the audio mixes to be listened to by the participants in our study. 
We did that by randomly selecting 45 seconds of every track, creating 3-minute long mixes.
The list of tracks used in the listening sessions,\footnote{\url{https://github.com/LPorcaro/longterm-musdiv/tree/main/material/listening_sessions}} and the audio mixes \footnote{\url{https://drive.google.com/drive/folders/1KLhBl5tXKeOUPzUzkG7JnqUdKi9e9woW}} are publicly available.

\section{Results}\label{05_results}
This section introduces the results of the analysis of participants' feedback and listening logs collected during the twelve weeks of the experiment. 
We start by describing in Section \ref{05_partstats} the population of our study, commenting on demographics, listening habits, and familiarity with EM, including in Section \ref{05_loganal} the analysis of the data retrieved from ListenBrainz. 
Then, we continue in Section \ref{05_grouping} reporting on the participants' group assignments, participation and drop-out rate. 
In Section \ref{05_lsanalysis}, we analyze participants' feedback during the 20 listening sessions.
Afterwards, Section \ref{05_EMFanal} includes the longitudinal analysis of the Electronic Music Feedback (EMF) questionnaire's responses, focusing first on openness and implicit association, and then on EM stereotypes.
Lastly, Section \ref{05_eos} presents the results of the End-of-Study survey.

\subsection{Participants' Demographics and Listening Habits}\label{05_partstats}
The exploratory nature of the study, and the lack of any meta-analysis that defines the ground truths of our variables, made it difficult to estimate with a power analysis the exact number of participants needed in order to observe potentially existing statistical differences. 
Nonetheless, guidelines from Human-Computer Interaction research helped us in estimating a valid number of participants \citep{Cohen1992, Kaptein2012, Kelly2016}. 
Indeed, when performing a t-test (or an equivalent non-parametric statistical hypothesis test) for the difference between two independent means, to observe a medium effect size (an effect likely to be visible to the naked eye of a careful observer), at a significance level of $\alpha=.10$ (acceptable for an exploratory study), with a statistical power of .80 (to avoid incurring a too great risk of a Type II error), according to Cohen \cite{Cohen1992}, it is required a sample of size 100 participants, i.e. 50 participants for each group. 
This is the reason why following the prescreening we recruited \textbf{110 participants} for this study, also foreseeing the eventuality of some of them dropping out.

Most of the selected participants are aged between 18 and 32 years old (91\%), come from Portugal (61\%), Italy (32\%) or Spain (7\%), almost equally divided into binary genders (53\% women and 44\% men)\footnote{Women and men include both cisgender and transgender identities.} with a small fraction of non-binary participants (3\%). 
In terms of education, 69\% have a bachelor's degree or lower, and 63\% indicate to still studying. 
According to their self-declared answers, the 88\% affirm having varied listening habits, only a quarter affirms listening often to EM, and a third affirms listening to a varied selection of EM. 
In terms of listening time, 78\% declare to daily listening on average between 1 and 3 hours of music.

The two familiarity tests (artist and genre, see Section \ref{03_studydesign}) help us in estimating the participants' knowledge of the Electronic Music (EM) scene. 
The test score ranges from 0, if no item in the lists is known, to 10.5 if all items are known.
In the artist test, the average score obtained by the selected participants ($N=110$) is 1.7$\pm$1.2. 
Instead, the average score of all the participants in the prescreening ($N=437$) is 2.6$\pm$2.1. 
In terms of genre familiarity, the average score is 5.1$\pm$1.6 against 5.7$\pm$1.9. 
Averaging over the two tests, the recruited participants got 3.4$\pm$1.1 against 4.2$\pm$1.8. 
As these numbers evidence, the filter applied during the prescreening made it possible to select a group of participants on average less familiar with mainstream EM artists and genres in comparison to a wider group of listeners with shared demographics.

Further characterization of the study participants may be done by looking at the average d-score (implicit association) and o-score (openness) obtained after the first Electronic Music Feedback (EMF) questionnaire in the \textit{PRE} stage. 
Indeed, at the beginning of the experiment participants affirmed being open to listening to Electronic Music, with a median o-score of 4 with an interquartile range of 2.75, and did not attach a negative nor positive valence to EM, having an average d-score of --0.09$\pm$0.42. 
In the next section, we further characterize the study participants by analyzing their listening logs.

\subsubsection{Listening Logs Analysis}\label{05_loganal}
The ListenBrainz listening logs give us an alternative perspective for understanding the relationship that the participants have with EM. 
Unfortunately, the analysis comprehends only data from 66 accounts because technical issues limited the stability of the connection between Spotify and Listenbrainz, which eventually made it impossible to collect data from all the participants. 
Nonetheless, the collected data are representative of some trends that we describe hereafter.

\begin{figure}[h!]
\centering
\includegraphics[width=0.8\textwidth]{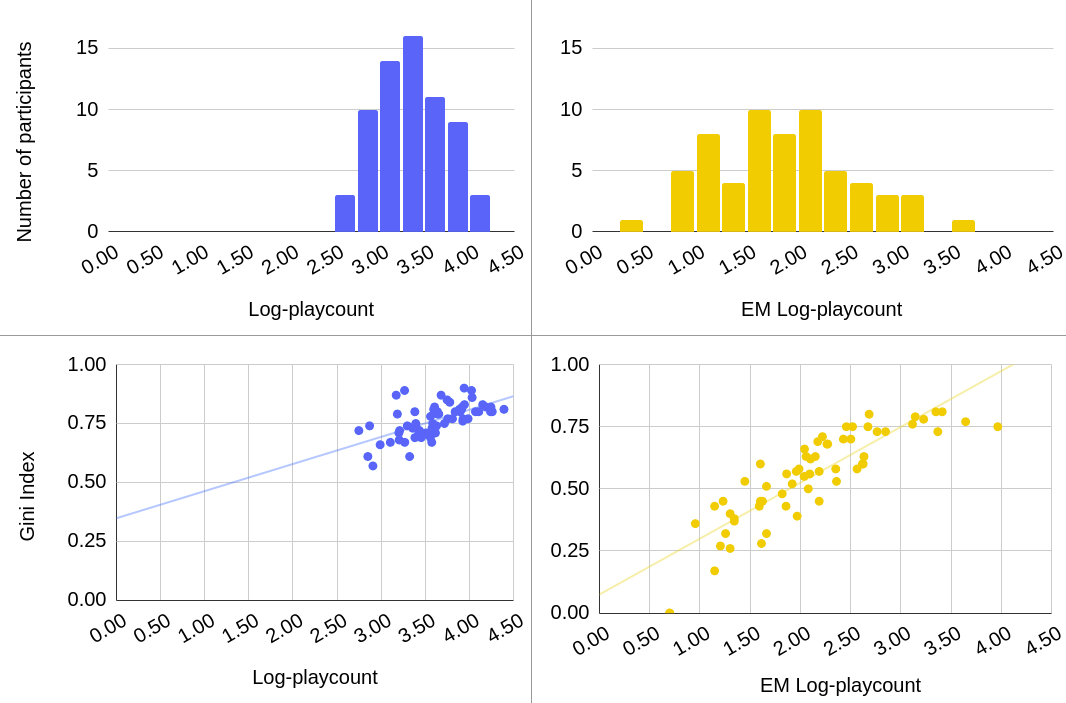}
\caption{Distribution of the participants' log-playcount (\textit{top}), and log-playcount versus Gini index (\textit{bottom}), computed with the whole set of logs (\textit{left}), and only with EM logs (\textit{right}).}
\label{fig:5_11}
\end{figure}

We start by analysing the participants' log-playcount over the course of the study, considering first the whole set of listening logs, and then only the EM logs.\footnote{We consider as \textit{EM log} a track listened to by a participant that is composed or performed by an artist associated with one or more EM genres among the ones retrieved from the Spotify API.}
On average, participants listened to 1479 tracks over the course of the study, more or less a daily hour of music if considering 3-minute long tracks.
In contrast, during the study they listened on average only to 56 EM tracks, meaning more or less 1 EM track a day. 
Figure \ref{fig:5_11} (\textit{top}) displays the distribution of the log-playcount separately over the whole tracks (\textit{left}), and only for EM tracks (\textit{right}).
We may observe that several participants have an EM log-playcount lower than 1, i.e. they listened to less than 10 EM tracks over the course of the study. 
Overall, for more than half of the study participants EM represents less than 15\% of the whole music listened to.

Moreover, to understand the variety of their listening habits, we compute the Gini index separately with the whole set of listening logs, and only with the EM logs.
For the latter set, the average Gini index (0.43$\pm$0.20) is smaller than the one computed with the whole set of logs (0.62$\pm$0.09).
This indicates that participants seem to have on average more varied habits in terms of EM in comparison to the whole music they listened to.
Figure \ref{fig:5_11} (\textit{bottom}) shows the relationship between the Gini index and log-playcount.
We observe that they are positively correlated ($\rho=.62$, $p<.01$), with an even stronger linear correlation between the two variables in the case of EM logs ($\rho=.86$, $p<.01$).
This supports the idea that the more a participant listened to EM, the less varied it was.
Hence, also taking into account the self-declared and estimated not-expertise of participants with respect to EM, we hypothesise the presence of two groups of listeners in our study: 
a) occasional heterogenous EM listeners, with a low Gini index (0.0--0.5) and low log-playcount (0--2); 
b) more-than-occasional homogenous EM listeners, with a high Gini index (0.5--1.0) and high log-playcount (2--4).

What quantitatively shown until now give us an idea of the study participants' listening habits, which however can be further complemented by looking at what kind of Electronic Music was most listened to.
Figures \ref{fig:5_13}, \ref{fig:5_15} in Appendix B display the top genres and artists ranked by popularity in the participants' logs, from which we may infer some trends.
Indeed, we note that \textit{House} and \textit{Electronic Dance Music (EDM)} alone constitute 75\% of the EM that participants listened to over the twelve weeks.
Such results do not come as surprise, because as previously commented the demographic segmentation of the participants in our study to some extent is similar to the one of \textit{EDM} listeners \citep{Voigt2016}. 
The most frequent artists in the logs, \textit{David Guetta}, \textit{Avicii} and \textit{Alok}, among the most popular in the scene, confirm the predisposition of the participants to mainstream EM.

Based on these observations, we may draw the following picture of the population of our study. 
They are Millenials and Gen Z from Southern Europe, equally divided into binary genders with a small fraction of non-binary, mostly without a graduate degree and still studying. 
Average in terms of time spent listening to music and having a self-appearance of having heterogeneous preferences, they affirm to not be heavy listeners of EM nor particularly varied in terms of EM listened to. 
Familiar with the mainstream EM artists and genres, but far to be considered experts of the genre, they may be grouped into occasional heterogenous EM listeners and more-than-occasional homogenous EM listeners, mostly listening to \textit{House} and \textit{EDM}. 
%

\subsection{Grouping, Participation and Dropout Rate}\label{05_grouping}
After selecting the participants matching our prescreening criteria, we needed to split them into two groups, one to be exposed to recommendations with high diversity and one with low diversity. 
During the pilot study we realised that, given the size of the sample, randomly assigning participants to groups could result in an unbalanced baseline for the variables we were interested in studying. 
For instance, one group could have been formed by participants much more open to listening to EM than the other group. 
In order to avoid imbalance between characteristics, which may have affected the impact of the recommendations, we assigned participants using \textit{covariate adaptive randomization} \citep{Suresh2011}, creating two groups balanced in terms of familiarity with EM (\textit{familiarity test score}), openness in listening to EM (\textit{o-score}), implicit association with EM (\textit{d-score}), and the number of EM tracks listened to during the \textit{PRE} stage. 
At the end of this process, we got a group of 55 participants assigned to the High Diversity (HD) condition, and a group of 55 participants assigned to the Low Diversity (LD) condition, from now on the HD group and the LD group.

Table \ref{tab:05_part} summarises the participation of the two groups in the EMF questionnaire and in the listening sessions. 
Not surprisingly, participation decreased over the course of the study, notably between the \textit{PRE} and \textit{POST} phase in the EMF questionnaire (-19\%), and less between the first and fourth week of the listening sessions (-4\%). 
Overall, we may notice a small difference between the two groups, with the LD participants being more active during the study than their HD counterparts. 
Some participants have been excluded during the course of the study if
a) they never showed up after being selected (\textit{initial nonresponse}), or 
b) after a certain point they stop participating (\textit{attrition}). 
The initial nonresponse was quite high for the EMF questionnaire (HD: 7\%, LD: 5\%), but relatively small for the listening sessions (HD: 5\%, LD: 0\%). 
Instead, attrition was equal for both tasks (HD: 7\%, LD: 2\%). 
These numbers are in line with retention rates observed in Prolific and generally in longitudinal studies \citep{Leeuw2015, Kothe2019}.

In conclusion, we excluded from the analysis the responses of the participants who did not participate in: 
a) more than 4 listening sessions, and b) more than 3 EMF questionnaires. 
With this choice, we ended up analysing the responses of 94 over 110 participants (85\%), 45 over 55 in the HD group (82\%) and 49 over 55 in the LD group (89\%). 
\begin{table}[h!]
\centering
  \caption{Summary of the participation in the study. LS values are the median participation over each week.}
\scalebox{1}{
  \begin{tabular}{lrrrrrr}
    \toprule
    &\makecell[c]{PRE}&\makecell[c]{COND1}&\makecell[c]{COND2}&\makecell[c]{COND3}&\makecell[c]{COND4}&\makecell[c]{POST}\\
    \midrule
    \multicolumn{7}{l}{\textit{Electronic Music Feedback (EMF)}} \\
    HD &55 (100\%)&47 (85\%)&46 (83\%)&44 (80\%)&46 (83\%)&44 (80\%) \\
    LD &55 (100\%)&50 (90\%)&49 (89\%)&47 (85\%)&50 (90\%)&46 (83\%) \\
    \textbf{All} &\textbf{110 (100\%)}&\textbf{97 (88\%)}&\textbf{95 (86\%)}&\textbf{91 (82\%)}&\textbf{96 (87\%)}&\textbf{90 (81\%)}\\
    \midrule
    \multicolumn{7}{l}{\textit{Listening Sessions (LSs)}} \\
    HD&\makecell[c]{---}&50 (90\%)&47 (85\%)&48 (87\%)&48 (87\%)&\makecell[c]{---}\\
    LD&\makecell[c]{---}&52 (95\%)&50 (90\%)&52 (95\%)&49 (89\%)&\makecell[c]{---}\\
    \textbf{All}&\makecell[c]{---}&\textbf{102 (92\%})&\textbf{97 (88\%)}&\textbf{100 (90\%)}&\textbf{96 (88\%)}&\makecell[c]{---}\\
    \bottomrule
  \end{tabular}}
  \label{tab:05_part}
\end{table}

\subsection{Listening Sessions Analysis}\label{05_lsanalysis}
During the 20 listening sessions in the \textit{COND} stage, we collected four types of data to measure the impact that recommendations had on the participants of the High Diversity (HD) and Low Diversity (LD) groups:
\begin{itemize}
    \item[--]\textit{Playlist accesses}: assigning 1 to a participant who after a session chose to explore the session's playlist, 0 otherwise.
    \item[--]\textit{Playlist interactions}: YouTube playlist's view count, aggregated over each group of participants. 
    \item[--]\textit{Like ratings}: ranging from -2 if a listening session was totally disliked by a participant, to 2 if a session was totally liked. 
    \item[--]\textit{Familiarity ratings}: 1 if the tracks in a listening session were familiar to a participant, -1 if unfamiliar, 0 if unsure. 
\end{itemize}

\begin{figure}[b!]
\centering
\includegraphics[width=0.9\textwidth]{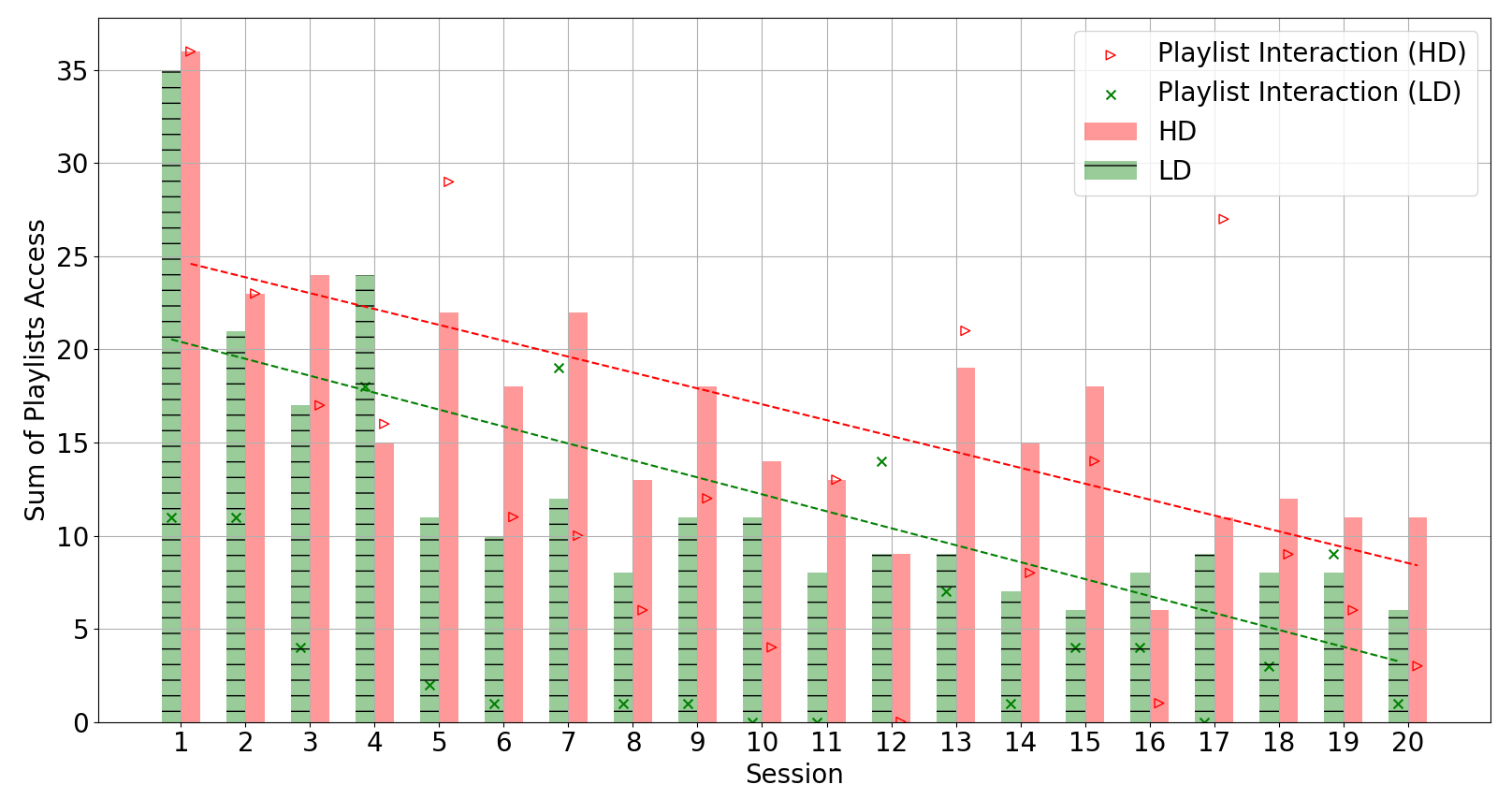}
\caption{Distribution of the playlists' accesses and interactions.}
\label{fig:5_16}
\end{figure}

Figure \ref{fig:5_16} displays the distribution of playlists' accesses and interactions, where trend lines are computed with the linear least squares method.
Similarly to the overall decrease in study participation, we may note that over the course of the sessions the overall engagement with the playlists decreased. 
Looking at the number of accesses and interactions, the HD participants seem to be more interested in discovering the music listened to than the LD ones. 
Moreover, we notice that on several sessions LD participants accessed a playlist but had zero interactions with it (e.g., sessions 10 and 11). 
This phenomenon never occurred to HD participants, who on the contrary in some sessions had much more interactions than accesses (e.g., sessions 5 and 17), meaning that some participants interacted several times with the same playlist. 
Naturally, in both groups accesses and interactions are positively correlated (HD: $\rho=.75$, LD: $\rho=.51$).

\begin{figure}[h!]
\centering
\includegraphics[width=0.9\textwidth]{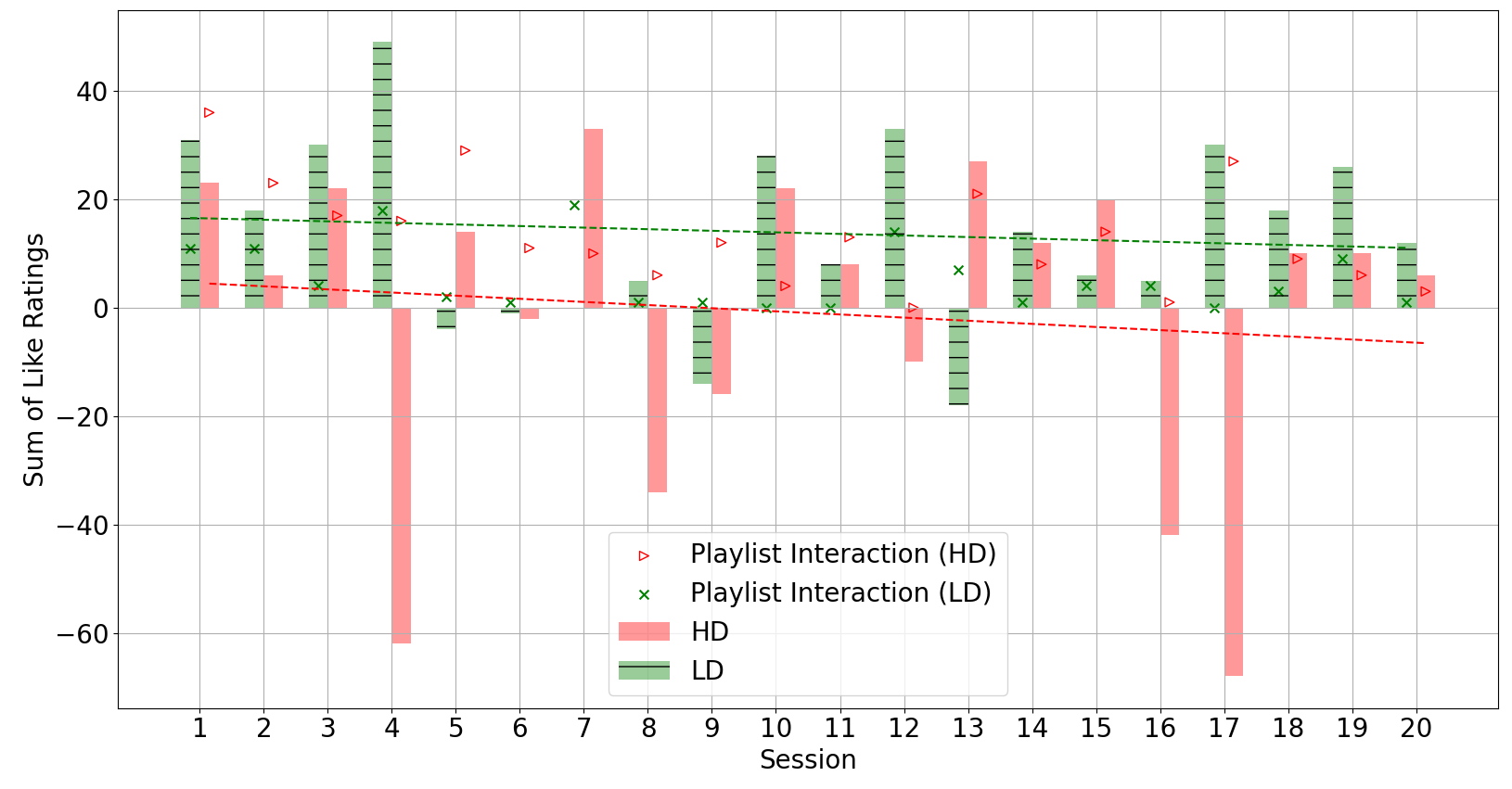}
\caption{Distribution of like ratings and playlists' interactions.}
\label{fig:5_17}
\end{figure}

\begin{figure}[h!]
\centering
\includegraphics[width=0.9\textwidth]{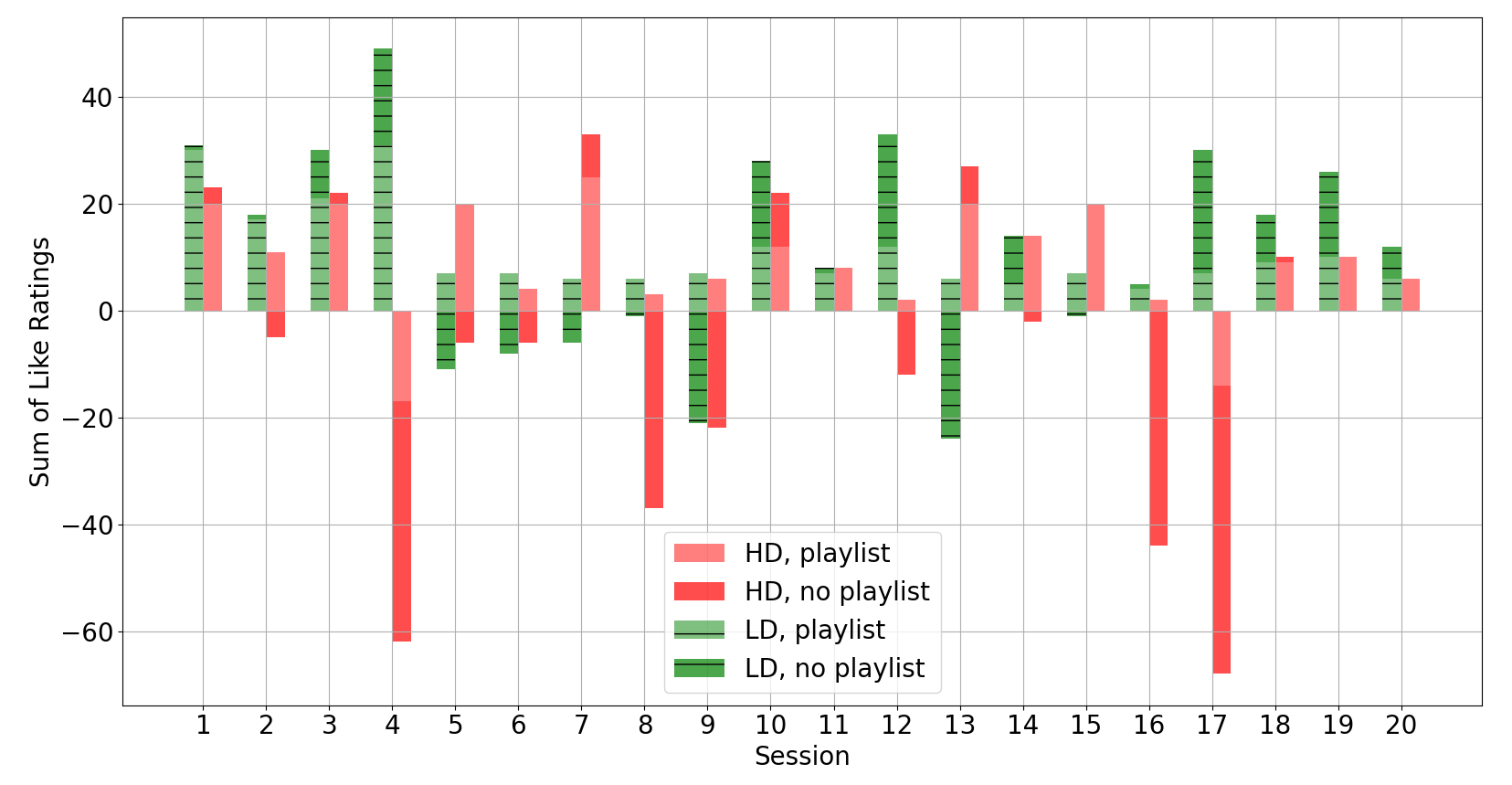}
\caption{Distribution of like ratings split among participants who accessed the playlist (\textit{light bar}) and those who did not (\textit{dark bar}).}
\label{fig:5_18}
\end{figure}

Figures \ref{fig:5_17} and \ref{fig:5_18} merge the playlists' data and the like ratings, giving us an alternative view for understanding the participants' reception of the recommendations. 
In Figure \ref{fig:5_17}, it emerges that the HD group disliked more sessions and with more extreme ratings compared to the LD group, who on average seem to have appreciated most of the music they have been exposed to. 
Nevertheless, HD participants even when they mostly disliked a session, interact with the playlists on YouTube (e.g., sessions 4 and 17).

Such behaviour is confirmed in Figure \ref{fig:5_18}, where like ratings are split between participants who accessed the playlists (\textit{light bars}) and those who did not (\textit{dark bars}). 
We see that some of the HD participants choose to access the playlist and discover more about the tracks listened to even in the most disliked sessions (\textit{negative light bars}), a phenomenon not visible for the LD group. 
On the contrary, some of the LD participants even if liking the tracks listened to, choose to not interact with the playlists (\textit{positive bars with zero interactions}, e.g., Figure \ref{fig:5_17}, sessions 10 and 11), or not access the playlists (\textit{positive dark bars}, e.g., Figure \ref{fig:5_18}, sessions 4 and 10).

The familiarity ratings' results are shown in Appendix \ref{B_AddTabFig} (Figures \ref{fig:5_19} and \ref{fig:5_20}) and summarised as follows. 
The sum of ratings is negative for almost every session, indicating that the tracks were mostly unfamiliar to the participants.\footnote{There is one exception in session 4 for the LD group, presenting a positive peak not in line with the rest of the sessions. This is due to one of the tracks included in that session ``Around The World (La La La La) (Ultra Flirt Hands Up Remix Edit)'', a remix of the famous track by \textit{A Touch Of Class (ATC)}.}
This trend was expected because of participants' not-familiarity with EM, but also because of the popularity filter applied before creating the recommendations. 
Besides, we observe a positive correlation between like and familiarity ratings (HD: $\rho=.55$, LD: $\rho=.72$), confirming what previous music psychology scholars have extensively proven: the more familiar a track sounds, the more is likely to be liked.

These few observations led us to formulate and test the following hypothesis:
\begin{itemize}
    \item[\textbf{H1.}] The HD group will have more accesses to playlists than the LD group.
    \item[\textbf{H2.}] The HD group will have more interactions with playlists than the LD group.
    \item[\textbf{H3.}] The LD group will like the tracks more than the HD group.
    \item[\textbf{H4.}]The LD group will like more tracks without accessing the playlist than the HD group.
    \item[\textbf{H5.}]The HD and LD groups will have the same level of familiarity with the listened tracks.
\end{itemize}
We test the aforementioned hypothesis by performing Mann-Whitney U tests, commonly used to compare the differences between two independent samples when the sample distributions are not normally distributed and the sample sizes are small. 
Table \ref{tab:05_mannwhy} reports the outcomes of the tests. 
\textbf{H1} and \textbf{H2} are confirmed by a significant statistical difference between HD and LD groups in terms of playlist accesses and interactions. 
In terms of like ratings, we see a smaller effect size with only 62\% of sessions having HD group ratings lower than the LD ones (\textbf{H3}). 
This proportion increases to 70\% if looking only at the ratings of participants who did not access the playlists (\textbf{H4}).
This confirms that even when LD participants liked the listening sessions, they interacted with the playlists less than the HD group. 
Lastly, looking at the familiarity ratings no significant difference is found (\textbf{H5}), confirming the hypothesis and also that the design of the listening session was effective, exposing subjects to music they were mostly unfamiliar with. 

\begin{table}[h!]
\caption{Summary of the Mann-Whitney U test results. M stands for median value. CLES (Common Language Effect Size) is the proportion of pairs where x is higher than y. When the alternative is ``greater'' x are the values of the HD group and y of the LD group. With the value ``lower'' we have the opposite scenario.}
\centering
\scalebox{1}{
  \begin{tabular}{clcccccc}
    \toprule
    Hyp.&Data&$M_{HD}$&$M_{LD}$&U&Altern.&p-val&CLES\\
    \midrule
    \textbf{H1}&Playlist access&15.0&9.0&307&greater&$<$.01&.77 \\
    \textbf{H2}&Playlist interaction&11.5&3.5&301&greater&$<$.01&.75 \\
    \textbf{H3}&Like ratings&9.0&13.0&152&lower&.10&.62 \\
    \textbf{H4}&Like ratings and playlist access&-1.0&1.0&120&lower&.01&.70 \\
    \textbf{H5}&Familiarity ratings&-21.5&-24.0&180&two-side&.60&.45 \\
    \bottomrule
  \end{tabular}}
  \label{tab:05_mannwhy}
\end{table}

\subsection{EMF Questionnaire Analysis}\label{05_EMFanal}
Through the analysis of the listening sessions, we have shown the distinct reactions of the two groups of participants when receiving the music recommendations.
Hereinafter, we continue by focusing on the impact of the exposure to EM recommendations, first, in terms of openness in listening and implicit association, and second, in terms of stereotypes that participants associate with this music genre.

\subsubsection{d-score and o-score}
The d-score measures the implicit association with EM, having negative values if a negative association is present and positive values in the opposite case.
The o-score measures the openness in listening to EM, ranging from 0 if a participant is not open, to 5 if a participant is truly open. 
We collected these scores six times during the longitudinal study, first at the beginning (\textit{PRE}), four times during the conditioning phase (\textit{COND} 1--4), and lastly at the end of the study after twelve weeks from the start (\textit{POST}).

\begin{figure}[h!]
\centering
\includegraphics[width=\textwidth]{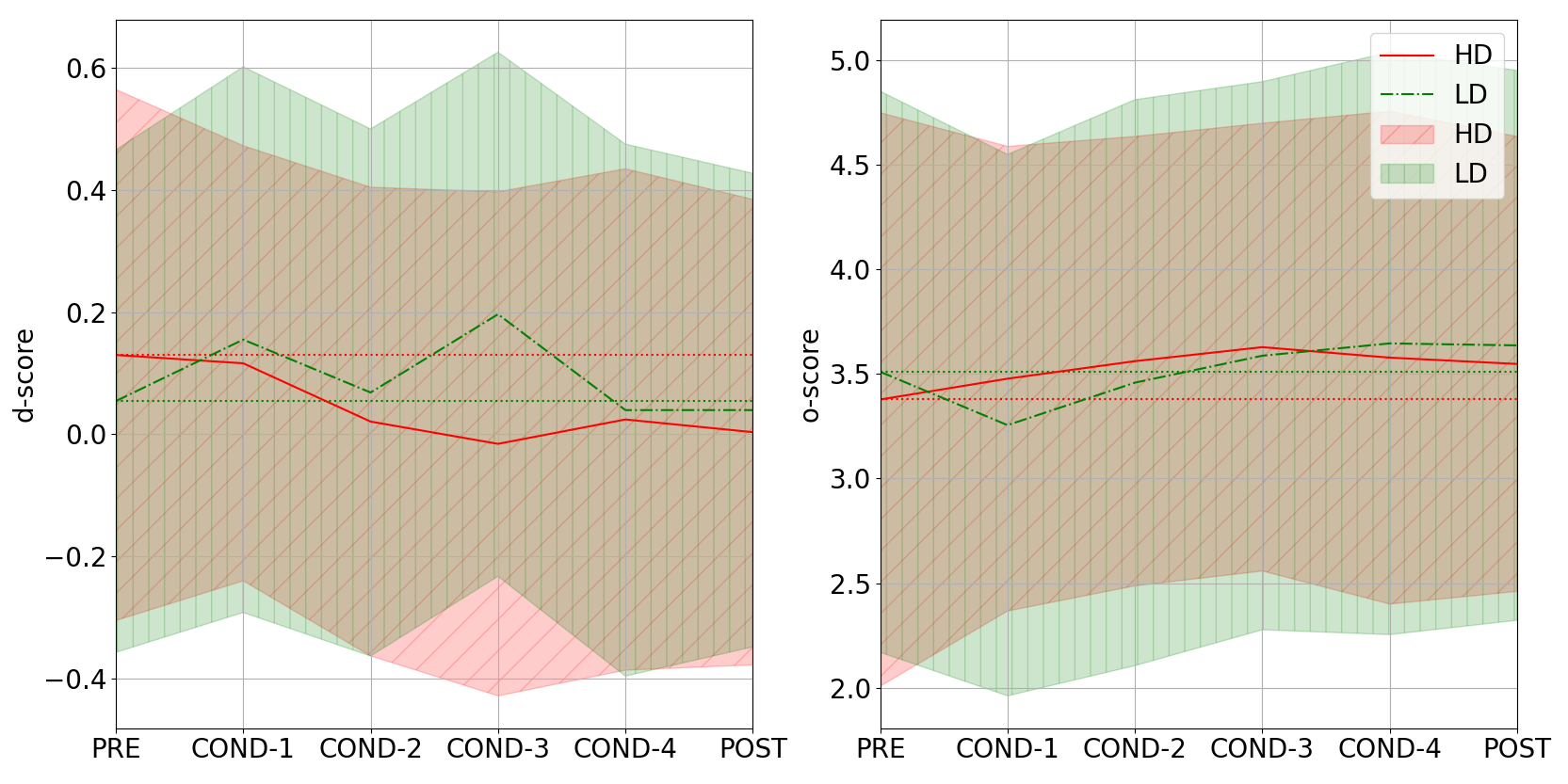}
\caption{Average and standard deviation of d-score (\textit{left}) and o-score (\textit{right}). The dotted lines are the average baseline measurements (\textit{PRE}). The filled area is the standard deviation from the mean value.}
\label{fig:5_21}
\end{figure}

Figure \ref{fig:5_21} shows the average scores and standard deviations separately for the two groups. 
In terms of d-score, we may observe that HD participants' scores starting from a positive average decrease toward zero.
Even if with more fluctuations, similarly the LD participants end up with an average score near zero, almost equal to the initial one. 
Instead, for the o-score both groups present a slight increase comparing the \textit{PRE} and \textit{POST} averages, with LD presenting a higher response variance.

Table \ref{tab:05_percatt} reports the percentages of participants grouped by their scores.
For the o-score, we split the participants into two groups, the less open in listening to EM having a score between 0 and 2, and the more open having between 3 and 5.
At the beginning (\textit{PRE}), the proportion for both HD and LD groups is around 75--25 (more open--less open), whilst in the \textit{POST} measurement the proportion of open participants slightly increases resulting in approximately 80--20. 

In terms of the d-score, the proportions in the two groups are initially quite different, with the HD group having more positive scores than the LD group.\footnote{Even if the d-score was included as a covariate while assigning participants to groups using covariate adaptive randomization, the higher initial nonresponse in the HD group caused such imbalance.}
Notwithstanding, over the course of the study the participants' scores move towards zero, with the neutral group (d-score $\in [-0.25, 0.25]$) consisting of almost half of the participants for both HD and LD groups, whilst the positive (d-score $>$ 0.25) and negative (d-score $<$ 0.25) scores are equally split. 

\begin{table}
\caption{Percentage of participants divided according to the scores collected.}
\centering
  \begin{tabular}{cc|ccc|ccc}
    \toprule
    &&\multicolumn{3}{c|}{\textbf{HD}}&\multicolumn{3}{c}{\textbf{LD}}\\
    &&PRE&POST&\textit{diff}&PRE&POST&\textit{diff}\\
    \midrule
    \textit{o-score}&0-2&26.7\%&14.3\%&--12.4\%&24.5\%&18.2\%&--6.3\%\\
    &3-5&73.3\%&85.7\%&+12.4\%&75.5\%&81.9\%&+6.3\%\\
    \midrule
    \textit{d-score}&$<$--0.25&17.8\%&28.6\%&+10.8\%&28.6\%&22.7\%&--5.9\%\\
    &$\pm$0.25&42.2\%&45.2\%&+3.0\%&38.8\%&52.3\%&+13.5\%\\
    &$>$0.25&40.0\%&26.2\%&--13.8\%&32.6\%&25.0\%&--7.6\%\\
    \bottomrule
  \end{tabular}

  \label{tab:05_percatt}
\end{table}

\begin{figure}[b!]
\centering
\includegraphics[width=\textwidth]{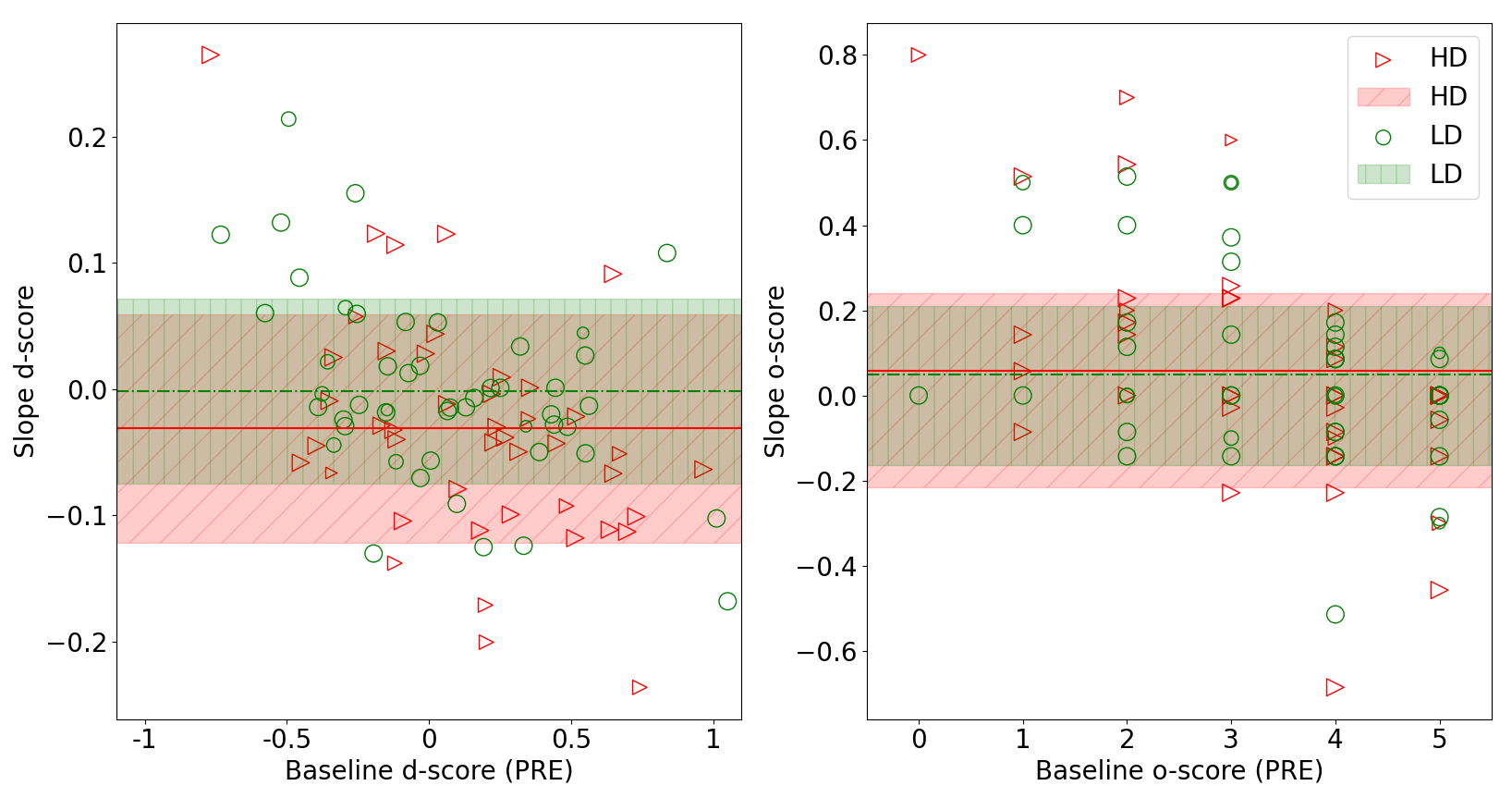}
\caption{Individual d-score (\textit{left}) and o-score (\textit{right}) slopes versus baseline (\textit{PRE}) scores. The horizontal line is the mean slope for each group, while the filled area represents the standard deviation.}
\label{fig:5_22}
\end{figure}

This analysis shows few aspects of the average behaviour of the HD and LD groups, without however considering individual differences. 
In order to further confirm what found at the group level, we explore the association between the rate of change and the initial scores by using the individual slopes obtained from the regression analysis of each participant's scores.
Figure \ref{fig:5_22} shows the slopes describing the trajectory of each participant versus the baseline scores obtained at the beginning of the study, separately for the d-score and o-score. 
Each point in the scatter represents a participant, where if the slope is positive indicates that through the twelve weeks her scores increased, if negative decreased.

The slopes of the d-score are mostly clustered around zero, meaning that the implicit association towards EM did not extremely change for most of the participants. 
In the bottom-right quadrant, a greater presence of HD participants is visible, who represent the subjects starting with a positive attitude and then moving towards more negative ones. 
On the contrary, in the top-left we see mostly LD participants, representing the opposite scenario.
The average slope for the LD group is almost zero, whilst for the HD group is negative, confirming that, overall, participants of this last group developed less positive implicit association during the study.

We observe a different situation for the o-score, where no particular differences are observed between groups.
Only in the case of the two participants who started the experiment declaring to be not open to listening to EM neither for one hour a month (baseline score equal to zero), we clearly observe different slopes.
Indeed, the LD participant over the weeks seems to have not changed her openness, instead, the HD participant has a positive slope. 
This indicates that even if at the beginning she was not open to listening to EM at all, eventually she started to be more open during the study.

Starting from these observations, we formulate two hypotheses on the impact of recommendations on the scores. In fact, over the course of the study for the whole group of participants we hypothesise that: \textbf{(H6)} the implicit association with EM will tend towards neutral valence, and \textbf{(H7)} the openness in listening to EM will increase. 
We use the Wilcoxon signed-rank test for testing this hypothesis. 
Two comparisons are made, first between the scores in the \textit{PRE} stage and the ones at the end of the fourth week of the \textit{COND} stage (\textit{PRE-COND}), and then between \textit{PRE} and \textit{POST} stage (\textit{PRE-POST}). 
Using the former, we are able to measure the impact of recommendations on participants right after being exposed to EM, while with the latter we measure if the impact still exists after one month from the exposure.

Table \ref{tab:wilc_scores} reports the outcomes of the tests.
In the case of the d-score, after the exposure participants' scores tendentially decrease, a trend confirmed when looking at the differences between the beginning and the end of the study. 
Instead, by analysing the o-score we observe an opposite behaviour, having an increase right after the exposure, which becomes not significant comparing the \textit{PRE} and \textit{POST} measurements. 
However, for both scores the effect size was not particularly large.
As a further step, by means of correlation analysis we measure the temporal stability of the two scores considering again the two intervals \textit{PRE-COND} and \textit{PRE-POST}. 
In terms of d-score we observe lower stability over time in comparison to the o-score both in the \textit{PRE-COND} measurements (d-score: $\rho=.30$, $p<.01$, o-score: $\rho=.57$, $p<.01$) and in the \textit{PRE-POST} measurements (d-score: $\rho=.34$, $p<.01$, o-score: $.53$, $p<.01$). 
These results corroborate the idea that implicit measurement may be less resistant to situationally induced changes than explicit measures \cite{Gawronski2017}.

\begin{table}[h!]
\caption{Summary of the Wilcoxon signed-rank test results. M stands for median value (1: \textit{PRE}, 2: \textit{COND} or \textit{POST}). CLES (Common Language Effect Size) is the proportion of pairs where x is higher than y.}
\centering
\scalebox{1}{
  \begin{tabular}{clcccccc}
    \toprule
    Hyp.&Data&$M_{1}$&$M_{2}$&W&Alt.&p-val&CLES\\
    \midrule
    \textbf{H6}&d-score (\textit{PRE}-\textit{COND})&.065&.008&2519.5&$>$&.01&.53 \\
    \textbf{--}&d-score (\textit{PRE}-\textit{POST})&.069&.016&2260.0&$>$&.05&.55 \\
    \textbf{H7}&o-score (\textit{PRE}-\textit{COND})&4.0&4.0&414.5&$<$&.05&.54 \\
    \textbf{--}&o-score (\textit{PRE}-\textit{POST})&4.0&4.0&434.5&$<$&.16&.52 \\
    \bottomrule
  \end{tabular}}
  \label{tab:wilc_scores}
\end{table}

After highlighting the overall impact of recommendations on the study participants, we are interested in understanding the role of diversity in such change. 
Therefore, we implement two methods to perform a \textit{PRE-POST} analysis, described as follows. 
We denote $X_i = 0$ if a participant is part of the LD group, and $X_i = 1$ if part of the HD group. 
For each participant, $Y_{i0}$ is the measurement in the \textit{PRE} stage, and $Y_{i1}$ at the \textit{POST} stage.
We use two regression methods to compare the groups. 
In the \textit{follow-up analysis} we look at the difference in the mean response at follow-up (\textit{POST}) comparing the two groups: $Y_{i1}= \beta_0+\beta_1X_i+\epsilon_i$. 
Instead, in the \textit{change analysis} we study the difference between the average change (\textit{PRE-POST}) comparing the two groups: $(Y_{i1}-Y_{i0})= \beta_0+\beta_1X_i+\epsilon_i$. 
From the follow-up analysis, we have no evidence of a significant difference in the mean responses between HD and LD groups at the \textit{POST} stage, both for the d-score ($\beta_1=-.04$, $SE=.08$, $p=.66$) and the o-score ($\beta_1=-.09$, $SE=.26$, $p=.73$). 
Similarly, the change analysis does not evidence differences between groups in the average change between the beginning and the end of the experiment, both for the d-score ($\beta_1=-.09$, $SE=.10$, $p=.38$) and the o-score ($\beta_1=.1$, $SE=.27$, $p=.71$).

In summary, after the exposure to four weeks of music recommendations we found a slight change in implicit association (\textbf{H6}) and openness (\textbf{H7}), but we have not evidenced any particular influence by the degree of diversity at which participants were exposed. 

\subsubsection{Stereotype Analysis}
The results of this section of the EMF questionnaire are displayed in Figure \ref{fig:5_24}, \ref{fig:5_25}, and \ref{fig:5_26} (Appendix \ref{B_AddTabFig}) respectively for the listening contexts, the musical properties and the artists' characteristics that participants associated with Electronic Music (EM).
Hereafter, we summarise the main results. 
As done for the d-score and the o-score, we compare exclusively the measurements taken at the beginning of the study (\textit{PRE}), at the end of the listening sessions (\textit{COND}), and at the end of the study (\textit{POST}).

Among the eight contexts presented in the survey, participants indicate that they would preferentially listen to EM while doing a dynamic and energetic activity (\textit{partying}, \textit{running}, \textit{commuting}, and \textit{shopping}). 
On the contrary, they disagree that EM is suitable for being listened to during activities that require a higher level of calm or concentration (\textit{sleeping}, \textit{studying}, and \textit{relaxing} or \textit{working}).\footnote{There is some ambiguity about the definition of working which may have influenced the responses. Indeed, depending upon the type of work, it may be considered an energetic or a calm activity.} 
Nevertheless, the exposure to recommendations did not largely affect the opinion of the participants. 
Indeed, performing a Wilcoxon signed-rank test between \textit{PRE}-\textit{COND} and \textit{PRE}-\textit{POST}, for six contexts out of eight no statistically significant difference ($p < 0.05$) have been found. 
Likewise, by using the Mann-Whitney U test we have not found significant differences between the HD and LD participants' responses, indicating that the level of diversity did not differently affect the participants.

The only two contexts wherein we found significant differences are \textit{running} and \textit{shopping}. 
In the former case, the LD group is strongly convinced about the use of EM for running, with the percentage of agreement passing from 62\% in \textit{PRE} to 76\% in \textit{COND} and \textit{POST}. 
In the HD group, we observe an opposite tendency, passing from an agreement of 73\% in \textit{PRE} to 66\% and then 68\% in \textit{COND} and \textit{POST}. 
These results support the idea that being exposed only to \textit{trance} music, a high-energy kind of music, may affect the listeners in associating EM with a high-energy kind of activity like running.
On the contrary, while exploring different facets of EM listeners may have realised that some genres are not fitted for being listened to while running. 
Instead, in the case of \textit{shopping} we see that both groups start disagreeing in the \textit{PRE} measurement (HD: 52\%, LD: 62\%), but then over the course of the study arrive at a more balanced situation between agreement, disagreement and neutral responses. 
Observing less extreme responses among participants is reasonable being shopping an activity not so dynamic, for instance in comparison to \textit{running}, neither so calm, as for instance \textit{studying}.

Similarly, the musical properties that participants associate with EM tracks have not been largely affected by the recommendations. 
Among the four selected features, participants changed their opinion only on the presence of acoustic instruments, especially the ones in the HD group. 
Indeed, for them we find a significant difference ($p = .01$, $CLES=.32$) both comparing \textit{PRE}-\textit{COND} and \textit{PRE}-\textit{POST} measurements. 
Moreover, the Mann-Whitney U test confirms the significant difference between the HD and the LD groups’ responses both in \textit{COND} and \textit{POST} measurements ($p = .04$, $CLES=.61$). 
Observing the distribution in Figure \ref{fig:5_25}, we may notice that at the beginning 79\% of HD participants disagree on the fact that EM had mostly acoustic instruments, while in the \textit{COND} and POST measurements only about 50\% disagree. 
On the contrary, the percentage for the LD group remains quite stable over the course of the three months. 
This is consistent with the fact that the LD group has been exposed only to \textit{trance} music, which rarely has parts with acoustic instruments.
On the contrary, HD participants listening to genres such as \textit{Electroacoustic} may have changed their idea about the acousticness of EM.

Lastly, analysing which characteristics participants associate with EM artists (Figure \ref{fig:5_26}), no statistically significant differences have been found between HD and LD groups. 
Only in terms of age, we find a difference between \textit{PRE}-\textit{COND} and \textit{PRE}-\textit{POST} measurements.

\subsection{End-of-Study Survey Analysis}\label{05_eos}
The analysis of the End-of-Study (EoS) survey reveals a few more qualitative insights that complement what is presented in the previous sections. 
First, we analyse participants' answers concerning the openness and appreciation of EM, before, during and after participating in the study. 
Then, we focus on a set of stereotypes to see how exposure to EM recommendations has affected the participants' opinions. 
We recall that the survey is formed by three main groups of items, the first asking about the experience of participants before the study, the second during, and the third after the study.

\begin{figure}[h!]
\centering
\includegraphics[width=\textwidth]{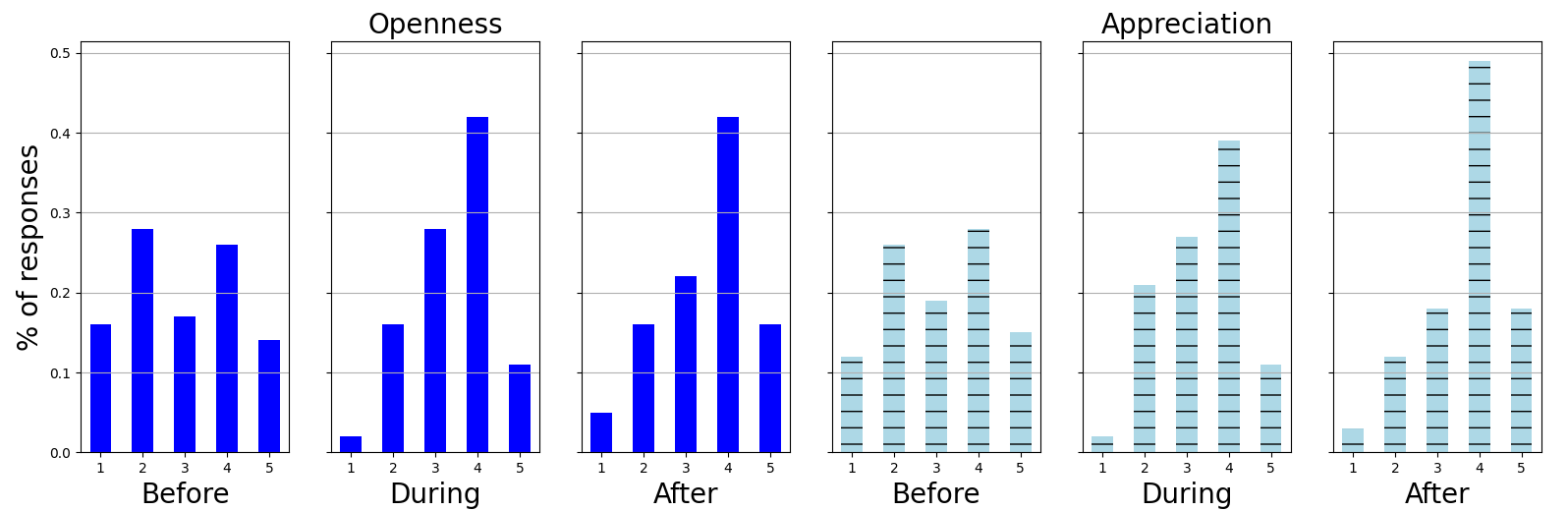}
\caption{Distribution of responses for the openness (\textit{dark blue}) and appreciation (\textit{light blue, dashed}) 5-point Likert scale. Response 1 corresponds to strong disagreement with the items, and 5 to strong agreement.}
\label{fig:5_27}
\end{figure}

The survey contains 16 Likert items to measure the participants' openness in listening to EM and 16 items for measuring appreciation of EM, divided into 6 items asking for participants' beliefs before, 6 during, and 4 after participating in the study. 
By analysing these three sets separately, we may get an idea about how participants perceived a change in their openness and appreciation of EM due to their participation in the study. 
Figure \ref{fig:5_27} presents the distribution of the responses for the whole group of participants. 
We avoid reporting results separately for the HD and LD groups because no significant difference has been found between their responses. 
Analysing the internal consistency of the sets of items computing Cronbach's alpha ($\alpha$), we observe an acceptable consistency in the three sets. 
Indeed, for the openness items we have: before $\alpha =.88$, during $\alpha=.82$, and after $\alpha=.72$.
For the appreciation items: before $\alpha=.90$, during $\alpha=.86$, and after $\alpha=.77$. 
Therefore, we may assume that the items properly reflect the two concepts that we wanted to measure.

Focusing on the openness in listening to EM, we notice that whilst participants' responses before the study were quite balanced around the neutral response, even if with high variance (M=$2.94\pm1.31$).
When asked about their openness during and after the study, they averagely declared to be more open in comparison with the beginning (respectively with $M=3.43\pm0.95$, and $M=3.48\pm1.08$).
Therefore, their perceptions are in line with the results of Section \ref{05_EMFanal} wherein an overall increase in openness was found by looking at the results obtained from the Guttman scale. 
In terms of appreciation, we notice a similar trend with an initial balanced situation at the beginning of the study ($M=3.07\pm1.27$), that increases towards more positive responses over the 12 weeks ($M=3.36\pm0.99$ and $M=3.67\pm1.01$, respectively during and after the study).

\begin{figure}[h!]
\centering
\includegraphics[width=\textwidth]{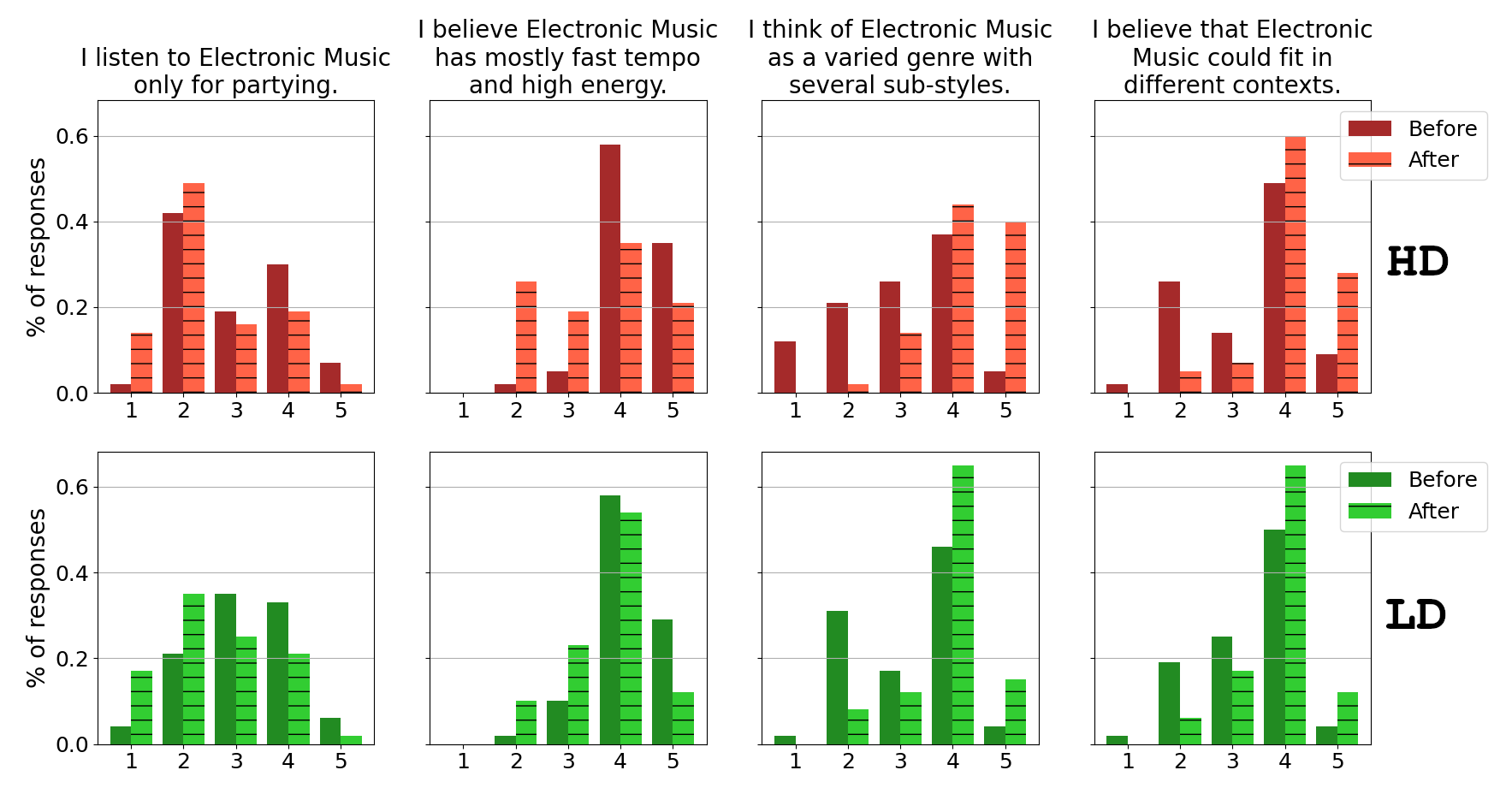}
\caption{Distribution of responses for the High Diversity (HD, \textit{top}) and Low Diversity (LD, \textit{bottom}) groups, before (\textit{clean bar}) and after (\textit{dashed ba}r) the participation in the study. Response 1 corresponds to “Totally Disagree” with the item, and 5 to “Totally Agree”.}
\label{fig:5_28}
\end{figure}

The second focus of our analysis is on comparing four items describing four stereotypes of EM, reported in Figure \ref{fig:5_28} together with the distribution of the responses. 
In this case, it displays the distribution for the HD and LD groups separately, because the impact of the recommendations on these items has been indeed mediated by their diversity.

For the first item analysed (``\textit{I listen to EM only for partying}''), we observe that prior to the study participants had on average a balanced response towards this stereotype (HD: $2.98\pm1.06$, LD: $3.17\pm0.97$).
However, during the study the two groups realise that may be restrictive to consider EM only for partying, with an overall decrease in their average response (HD: $2.47\pm1.03$, LD: $2.56\pm1.07$). 
In the case of the second item (``\textit{I believe Electronic Music has mostly fast tempo and high energy}''), we observe a similar behaviour.
In the beginning, participants strongly agree with this statement (HD: $4.26\pm 0.66$,  LD: $4.15\pm0.68$), but then after the study they changed their opinion (HD: $3.51\pm1.1$, LD: $3.69 \pm 0.83$), agreeing less strongly to the fact that EM has a mostly fast tempo and high energy.

The next considered item (``\textit{I think of Electronic Music as a varied genre with several sub-styles}'') presents an opposite situation in confront to the former two. 
Indeed, at the beginning on average participants neither agree nor disagree significantly with the statement (HD: $3.02\pm1.12$, LD: $3.19\pm1.0$).
However, participating in the study made them change their opinion about it, agreeing much more in percentage than before (HD: $4.21\pm0.77$, LD: $3.85\pm0.77$). 
Similarly, also in the case of the fourth item considered (``\textit{I believe that Electronic Music could fit in different contexts}''), we observe a neutral position of participants at the beginning (HD: $3.37\pm1.05$, LD: $3.35\pm0.9$), which later agree more with the fact that EM may fit in different contexts (HD: $4.12\pm0.73$, LD: $3.83\pm0.72$).
It is interesting to note that the responses in these two latter items are the only ones for which we found a significant difference between HD and LD subjects. 
Indeed, by performing a Mann-Whitney U test we obtain a p-value of .01 and .02 and a CLES of .65 and .62.  
This indicates that, even if for both groups the exposure to recommendation affected the beliefs about how varied EM may be and how it may fit in different contexts, in the case of the HD group this shift has been more pronounced.

\section{Discussion and Limitations}\label{06_discussion}
In this section, we discuss the results previously presented with the aim of defining, first, \textit{to what extent listeners' implicit and explicit attitudes towards an unfamiliar music genre can be affected by exposure to music recommendations?} (\textbf{RQ1}), and 
second, \textit{what is the relationship between music recommendation diversity and the impact on listeners' attitudes} (\textbf{RQ2)}.
We focus on four main aspects: impact on discovery (Section \ref{impact_disc}), implicit association  (Section \ref{impact_impl}), openness  (Section \ref{impact_open}), and stereotypes  (Section \ref{impact_stereo}).
Moreover, in each section we present the limitations and future work.

\subsection{Impact on Discovery}\label{impact_disc}
The most pronounced role that recommendation diversity plays in our study is with respect to the curiosity generated in the participants when exposed to music during the listening sessions.
Indeed, the listeners' willingness to explore Electronic Music (EM) playlists is significantly higher if exposed to highly diverse recommendations. 
Having in mind that EM was mostly unfamiliar to the subjects in our study, we believe that \textit{when listeners are not familiar with a music genre, a diversified set of recommendations could increase their willingness in exploring such music}. 
A message sent to us by a study participant supports this hypothesis:
``Hello! It was a pleasure participating, I discovered new artists that I really liked! Happy to have contributed to your study :) Thank you!''.

This is particularly important for the design of music recommender systems in streaming platforms, the digital places where most people land to discover new music \citep{Datta2018, Aguiar2018, Knees2019}. 
In offline settings, several studies have investigated the role that diversity may play, e.g \citep{Ferwerda2017}, but to our knowledge this is the first longitudinal user study that measures the long-term impact of recommendation diversity on music discovery. 
Recent works by Liang and Willemsen \cite{Liang2022a, Liang2022b} show that it is possible to favour exploration of distant music genres both in the short and long term by nudging listeners through specific design choices, for instance presenting such genres in the top of the recommendation lists. 
Starting from their findings, we believe that by mixing nudging mechanisms with diversification techniques practitioners may design recommendations that notably improve the experience of discovering unfamiliar music.

Nevertheless, the first limitation of this work rises from the definition of \textit{discovery} itself. 
Indeed, if as Nowak suggests discoveries are ``affective responses to music content that occur within individuals' life narratives and mediate their interpretation and definition of music'' \cite{Nowak2016}, we argue that the responses resulting from the participation in our study cannot be compared to what listeners usually experience in less artificial situations wherein they are exposed to music. 
Under a different lens, focusing on discovery strategies and behavioural attitudes Garcia-Gathright and colleagues \cite{Garcia2018} show how explorative goals may vary according to the listeners' needs.
A further step could be to link the various overt behaviours to people's affective responses, to evaluate if the discovery mediated by algorithmic recommendations has some shared values with other non-digital forms of curation (e.g., a DJ tracklist created for a live event).

\subsection{Impact on Implicit Association}\label{impact_impl}
The exposure to music recommendations helped participants in deconstructing part of the preexisting positive or negative association to EM, developing a more neutral attitude during the twelve weeks. 
The role of diversity here seems not significant, having the two groups of participants similarly behaved.
In fact, we argue that the tendency of the HD group to decrease more significantly in comparison to the LD group is because of the unbalance created by the different dropout rates.

Based on that, we hypothesise that \textit{when listeners are not familiar with a music genre, to receive repeated recommendations could mitigate the valence of the implicit association with such music}. 
The results obtained deviate from the findings in \cite{Vuoskoski2017}, where positive implicit attitudes towards facial images of people from two cultures, namely Indian and West African, are developed by mere exposure to music from such cultures. 
However, the types of association and stimuli that listeners experienced in our study are undoubtedly different. 

The following reasons may be at the root of the development of neutral attitudes towards EM in the presented study.
First and in line with the previous point on the discovery, the experimental setting mediated the affective response, also influencing the implicit association. 
The tendency of the participants to associate neutral valence with EM genres could be motivated by the artificiality of the events wherein participants listened to the music.
Second, given the participants' unfamiliarity with the genre, it is understandable that they did not develop any positive or negative attachment to genre labels that they might have never heard of before the study.

As elegantly discussed by McLeod \cite{McLeod2001}, genre namings are strictly tied to the EM communities behind which people identify, also influencing the dynamics of group formation. 
Therefore, we hypothesise that most of the genres shown in the SC-IAT test were only \textit{labels} to the study participants, and remained as such given that no mechanism of bounding was incentivized.
Indeed, no information about the genre of the tracks listened to was provided during or after the listening sessions. 
Under this lens, receiving neither positive nor negative responses from the participants may be seen as the desired result in the long term, which however needs further analysis to be understood in depth.

\subsection{Impact on Openness}\label{impact_open}
Whilst listeners' implicit association with EM became more moderate throughout the study, their openness to listening to EM increased. 
This indicates that in the long term exposure to unfamiliar music could favour listeners in being less reluctant in approaching a genre previously unknown. 
This finding does not come totally anew, partly confirming results from the literature on the impact of repeated exposure to music.

Diversity here seems to motivate in particular participants who started the experiment affirming to be not open in listening EM. 
Indeed, subjects who passed from not being open to being open in listening to EM for one hour or more a week are in the HD group twice the size of the LD group. 
Still, the impact of recommendation diversity is apparently not significant. 
Moreover, contrary to the implicit association, participants' openness was more consistently affected right after the conditioning stage than at the end of the study.
Connecting this with the impact on discovery, we deduce that openness is highly influenced by exposure, but such influence decays rapidly when listeners stop to be exposed to recommendations. 
In light of this, we support the idea that \textit{when listeners are not familiar with a music genre, repeated exposure to recommendations could increase their openness in listening to such music}. 
A message sent by one of the participants of the HD group right after the end of the \textit{COND} stage supports this intuition: ``Thank you so much [...] It was very interesting and a good opportunity to learn about this musical genre.''.

Again, the settings of the experiment may have influenced the outcomes responses.
In fact, participants were requested to be highly focused while listening to the music during the study, a condition which is not quite frequent in today's listening practices where music is often relegated to the background while performing other activities \cite{Morreale2021}. 
This could have affected the openness more consistently in comparison if tracks would have been passively listened to, for instance, while walking in a mall or while working. 
Repeating the experiment in a controlled non-online environment may lead to more precise results in this regard.

Furthermore, the openness measured using the Guttman scale has its own limitations. 
Indeed, listeners' openness in listening to EM could be determined by an overall curiosity in discovering music. 
For instance, a listener with very heterogeneous musical tastes could be open to listening every day to one hour of electronic music, one hour of classical music, one hour of rock music, and more. 
On the contrary, a very homogeneous listener would avoid listening to electronic, classical and rock music at the same time if disliked. 
We foresee the analysis of participants' openness in listening to EM with regard to their overall tendency of listening to varied music.

\subsection{Impact on Stereotypes}\label{impact_stereo}
What emerges from the analysis of the results is that the participants' idea of EM is quite stereotypical, and probably based exclusively on mainstream \textit{Electronic Dance Music} artists. 

Indeed, listeners not familiar with EM  may fall into the trap of misinterpreting it as music composed only with electronic instruments (e.g., drum machine, sampler, synthesizer). 
Instead, acoustic instruments, even if sampled, filtered, or generally modified, have always been used by EM artists. 
Under this lens, the fact that acousticness is the only feature on which participants’ changed their idea after the exposure to recommendations does not come as a surprise.  
Moreover, the stereotype that EM is only for parties, and the common belief that it has mostly a fast tempo and high energy, have been partly deconstructed by the exposure to music such as \textit{ambient}, \textit{electroacoustic} or \textit{chill out} respectively in sessions 2, 4 and 18 of the HD group.
Nevertheless, the characterization of EM as \textit{Energetic and Rhythmic} is quite common also in scientific literature, e.g., \cite{Rentfrow2007}, and participants of the LD group at some level experienced this aspect of EM. 
Besides, another accomplishment of this study is to show that listeners exposed to diversified recommendations may have a better understanding of the variety of EM culture, and realise that this music may fit in different contexts, not only for partying. 
Therefore, we deduce that \textit{exposing listeners not familiar with a music genre to a set of diversified recommendations, showing the different facets of such genre, could deconstruct pre-existing stereotypes}. 
This finding is in line with the music psychology literature on the influence of music exposure.

Likewise, the representation that participants have of EM artists is quite stereotypical, and also in this case it was something expected. 
EM artists are, according to them, mostly men, white, under 40 years old and coming from developed and high-income countries.
In this case, we did not expect that recommendations would affect participants’ opinions, mostly for two reasons which however are also two limitations of this work. 
First, participants' origin has been purposely restricted to a small part of the world, normally labelled as Western, Educated, Industrialised, Rich and Democratic (WEIRD) societies. 
Similarly, also the EM part of the study is somehow biased towards Western artists.  
In fact, the semi-automatic method for creating the dataset has produced recommendation lists that reflect the variety of EM in terms of different genres, but not its variety as music culture played in different parts of the world.

\begin{figure}[h!]
\centering
\includegraphics[width=\textwidth]{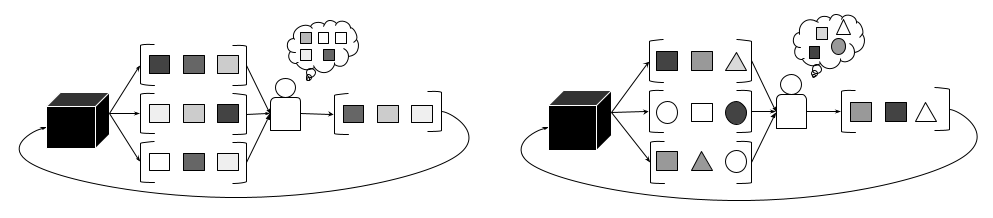}
\caption{A visual representation of the rationale of our study.}
\label{fig:5_23}
\end{figure}

In conclusion, Figure \ref{fig:5_23} depicts the rationale behind our study by using an analogy about how people create mental models or abstractions of a field, using as an example geometric shapes. 
If a geometric shapes recommender system exposes a user not familiar with shapes only to squares, even if she may interact with squares having different colours, her idea of geometric shapes will be centred around squares. 
Instead, a system recommending squares, circles and triangles may help her in learning about other geometric shapes, and eventually she will interact also with those, or maybe not. 
In the end, one of the functions of recommendation diversity may be to help users learn what different shapes are.

\section{Conclusions}\label{07_conclusions}
The impact assessment of music RS with regard to people's behaviour, attitudes, habits or beliefs, is an active and challenging research topic which is attracting more and more practitioners from different disciplines. 
Until now, simulation methods have been the most explored approach by the RS community to study the dynamics between users and items, and undoubtedly they have shed a light on several behavioural aspects of such interaction. 
Nevertheless, the fact that \textit{behaviourism is not enough} has already being pointed out by Ekstrand and Willemsen \cite{Ekstrand2016}. 
The focus only on what users do, listen or consume is indeed one of the main limitations we find in music RS research, which motivated us in designing this longitudinal study.

We did not target or expect to drastically change the listening habits of the study participants because the function of RS is not, and should not be, to manipulate listeners' towards specific artists or genres. 
Moreover, the socio-cultural background, empathetic and affective response, generational differences, music education and many more aspects through life define what people listen to, and with this regard, recommender systems are only one of the many ways thanks to which people discover and interact with music \citep{Born2021}. 
Nevertheless, the extensive use of music streaming services, especially by young generations, wherein algorithmic recommendations play a central role, raise questions in terms of human agency and autonomy of the next generation of music listeners \citep{Crider2022}, questions that RS practitioners should not neglect.

\textbf{Reproducibility}. To encourage reproducibility of the results, the code and data used in our study are publicly available.\footnote{TO BE READY IN CASE OF ACCEPTANCE: \url{https://github.com/LPorcaro/longterm-musdiv}}

\begin{acks}
This work is part of the project Musical AI - PID2019-111403GB-I00/AEI/ 10.13039/501100011033 funded by the Spanish Ministerio de Ciencia, Innovación y Universidades (MCIU) and the Agencia Estatal de Investigación (AEI).
This work is also partially supported by the HUMAINT programme (Human Behaviour and Machine Intelligence), Joint Research Centre, European Commission. 
The project leading to these results received funding ``la Caixa'' Foundation (ID 100010434), under agreement LCF/PR/PR16/51110009, an from 
EU-funded projects ``SoBigData++'' (grant agreement 871042) and ``FINDHR'' (grant agreement 101070212).

\end{acks}

\bibliographystyle{ACM-Reference-Format}
\bibliography{main}

\appendix
\newpage
\section{Diversification Design}\label{A_recdes}
This appendix outlines the diversification method designed to create the music recommendations for this study.
Section \ref{sec:05_dataset} describes the creation of the dataset, containing different kinds of Electronic Music (EM). 
After selecting a pool of approximately 1.5 thousand tracks from the dataset, several audio models were explored to understand what type of tracks' representation better served the purpose of diversifying the recommendations, as described in Section \ref{sec:05_models}. 
Lastly, we describe in Section \ref{sec:05_diversification} how starting from the tracks' embeddings we built the diversified recommendation lists to which study participants were exposed.

\subsection{Dataset}\label{sec:05_dataset}
One of the goals of the study was to expose people not familiar with the EM culture to different genres which could fit under this label.
Therefore, our aim was to select as many as possible varied tracks to represent the richness of this culture, even if we did not aim to create a dataset representing its entirety.
In order to do that, we designed a semi-automatic method based on two sources: \textit{Wikipedia} and \textit{Every Noise at Once}.\footnote{\url{https://everynoise.com}}

The initial step consisted in retrieving the list of EM genres from the Wikipedia page ``List of electronic music genres'' \cite{Wikipedia2022}.
Whilst several taxonomies and hierarchies of EM genres and subgenres have been published, we believe that the information on Wikipedia properly represents the variety of EM, enough precisely for our purposes. 
From that source, we obtained a set of \textbf{20 genres} and \textbf{320 subgenres}.\footnote{We refer to \textit{genre} when indicating the top-level genre labels in the Wikipedia hierarchy, highlighted in bold. Instead, we refer to the others as \textit{subgenres}.}
The second step was to map these subgenres to the ones part of the Every Noise at Once (ENO) website.
ENO is described by its creator Glenn Mcdonald as ``an ongoing attempt at an algorithmically-generated, readability-adjusted scatter-plot of the musical genre-space, based on data tracked and analysed for several genre-shaped distinctions by Spotify''\citep{Spotify2018}. 
Born as a debugging tool, in its current form the website presents for each genre label a playlist formed by approximately one hundred tracks, considered representative of the genre by Spotify algorithms. 
Mapping Wikipedia to ENO, we linked to the \textbf{20 genres} only \textbf{181 subgenres}, for which we retrieved the corresponding playlists.

It should be noted a few aspects of the aforementioned approach. 
First, even if we are aware of the dynamical, intrinsic, ambiguous, and context-dependent nature of concepts such as genre and style \cite{Porcaro2021}, approaching music cultures as broad as Electronic Music is quite natural to identify different aesthetic and social characteristics in its several subgenres \citep{McLeod2001}. 
For instance, even without knowing anything about EM genres, it is possible to recognize the differences between the fast breakbeats of the \textit{Drum and Bass} and the slow-tempo beats of \textit{Downtempo}.
As a matter of fact, in a previous study we showed how familiarity with styles and subgenres highly influence the perceived diversity of people exposed to EM tracks \cite{Porcaro2022b}. 
Even if it could have left out some genres, pursuing a bottom-down approach to find representative Electronic Music, starting from genre labels arriving at tracks, seems to us the most obvious approach to creating a varied EM dataset.

Second, the choice of using ENO playlists to select candidate tracks could be criticised because of the opaqueness of the Spotify algorithms. 
Indeed, it is not possible to find the exact description of the procedure which assigns a genre label to artists and albums. 
Even having in mind this limitation, after exploring in depth the ENO website we believe that it provides a good representation of the genres on its map, sharing the idea of ENO creator that
``the point of the map, as with the genres, is not to resolve disputes but to invite you to explore music'' \citep{McDonald2013}. 
Under this lens, we do not argue that the classification in this study is the ultimate one, but one of the possible classifications which may help listeners navigate the EM culture.

After a preliminary exploration of the dataset, we performed a manual cleaning of some of the music selected to be sure that crossover genres would not enter into the final listening sessions.
We do not argue if \textit{Electronic Rock} may be considered \textit{Electronic} music, \textit{Rock} music, or both, but we believe that crossover music did not fit the purpose of our study. 
In the end, we restricted the dataset to \textbf{20 genres} (\textit{ambient}, \textit{bass music}, \textit{breakbeat}, \textit{chill out}, \textit{disco}, \textit{drum and bass}, \textit{electroacoustic}, \textit{electronica}, \textit{garage}, \textit{hardcore}, \textit{hardstyle}, \textit{hauntology}, \textit{house}, \textit{IDM}, \textit{jungle}, \textit{noise}, \textit{plunderphonics}, \textit{techno}, \textit{trance}, \textit{videogame}), and \textbf{165 subgenres} listed in Appedix B (Table \ref{tab:05_genres}).
At that point, we had a pool of around 16 thousand tracks ($\sim$100 tracks for subgenre) listenable on Spotify.

Further filtering was applied by looking at the popularity of the tracks. 
Indeed, several studies have proven the influence of familiarity on music preferences \citep{Chmiel2017}. 
In order to avoid creating a \textit{popularity effect} in our study, i.e. participants' ratings influenced by the popularity of a track, we filtered the dataset by using two indicators: 
1) Spotify track' popularity; 2) YouTube view count. 
In detail, we implement the following procedure. 
From the 16 thousand tracks for which we already had the Spotify ID, we filtered out the ones with popularity less than the first quartile Q1 and major than the third quartile Q3 computed on the Spotify popularity indicator.
For the remaining ones, we made use of a Python wrapper to search with the YouTube API for the corresponding video by using the track names. 
As a result, we obtain a total of approximately 8 thousand tracks with Spotify ID and YouTube ID, already filtered by Spotify popularity.
The last step was to further filter out tracks according to the YouTube view count applying the same logic of Spotify popularity, including only tracks with views between Q1 and Q3. 
After the second filtering, we randomly choose ten tracks for each subgenre obtaining \textbf{1444 candidate tracks}, for which finally we extracted the audio embedding, as reported in the next section.

\subsection{Audio Models}\label{sec:05_models}
The use of deep representation models in MIR research is widespread and applied in several retrieval and classification tasks, e.g., auto-tagging, instrument recognition, genre classification, and ultimately music recommendations \citep{Won2022}. 
Nevertheless, the trustworthiness of such representations is still under the scrutiny of the research community, partly because of the still low interpretability in comparison to traditional hand-crafted music representations, usually informed by human music domain knowledge \citep{Kim2019}.
Even if the goal of this study was not to perform a rigorous comparative analysis of different music representation models, we were nevertheless interested in creating diversity-aware content-based music recommendations, on the one hand, based on state-of-the-art deep learning models, on the other hand enough interpretable to make us aware of the characteristics underlying the diversification outcomes. 
Consequently, we first started exploring several deep representation models, and then we attempted to interpret those by using a set of hand-crafted features.

\subsubsection{Deep Representation Models}
We experimented with four audio representation models available in \textit{Essentia} \citep{Bogdanov2013}, an open-source library for audio and music analysis. 
Notably, Essentia provides Tensorflow deep learning models built on different architectures and trained with different datasets, open sources and publicly available \citep{PabloAlonsoJimenez2020}. 
Among those available, we tested the following models:
\begin{itemize}
    \item \textbf{EffNet--Discogs}: EfficientNet \citep{Tan2019} trained with the Discogs-Effnet Dataset (DED).
    \item \textbf{MusicNN--MSD} / \textbf{MusicNN--MTT}: MusicNN \citep{pons2018atscale} trained with the Million Song Dataset (MSD) \citep{BertinMahieux2011} and the MagnaTagATune (MTT) dataset \citep{Law2009}. 
    \item \textbf{VGGish--AudioSet}: VGG \citep{Simonyan2015} trained with the AudioSet dataset \citep{Gemmeke2017AudioSA}. 
\end{itemize}
A detailed description of the models, datasets and implementations can be found online,\footnote{\url{https://essentia.upf.edu/models.html}} but here we discuss a few aspects relevant to our study. 
First, we employed these models as feature extractors, i.e. to obtain an embedded representation of each audio in our dataset. 
However, their intended purposes are quite different. 
For instance, \textit{VGG} was proposed to tackle the task of image recognition, while \textit{MusicNN} primarily focuses on auto-tagging. 
Therefore, our final choice was based not on the architecture which better performs according to its original scope, but on the feature extractor that better worked according to our objectives. 
Second, we intentionally selected a quite heterogeneous set of models, especially in terms of datasets used in the training stage. 
Indeed, \textit{AudioSet} is formed by almost two million clips annotated using sound labels not always specific to music (e.g. footstep, bark, cutlery). 
Instead, \textit{MSD} and \textit{MTT} datasets are annotated with 50 tags describing the genre, instrumentation or also mood of the tracks. 
Lastly, \textit{DED} contains tracks annotated with 400 music styles according to the taxonomy of the crowdsourced database \textit{Discogs}.

Once selected the models (\textit{EffNet--Discogs}, \textit{MusicNN--MSD}, M\textit{usicNN--MTT}, \textit{VGGish--AudioSet}), we extracted the audio embeddings for the 1444 candidate tracks part of our dataset. 
In the former three cases, we obtained 200-dimensional embeddings while from the \textit{VGGish-AudioSet} we get 1024-dimensional embeddings. 
At that point, we were interested in a model which could coherently represent Electronic Music according to the genre labels that we already got.  
Indeed, assuming that the tracks representative of one genre, i.e. coming from the same \textit{Every Noise at Once} playlist, should be more similar to one another in comparison to the tracks of another genre, our problem was translated into measuring how good the embeddings placed tracks from the same genre near in the embedded space. 
Therefore, we used the tracks' genre labels to create 20 clusters, and then we measured the consistency of each cluster by performing a Silhouette analysis \citep{ROUSSEEUW198753}.

\begin{figure}[t!]
\centering
\includegraphics[width=0.99\textwidth, ]{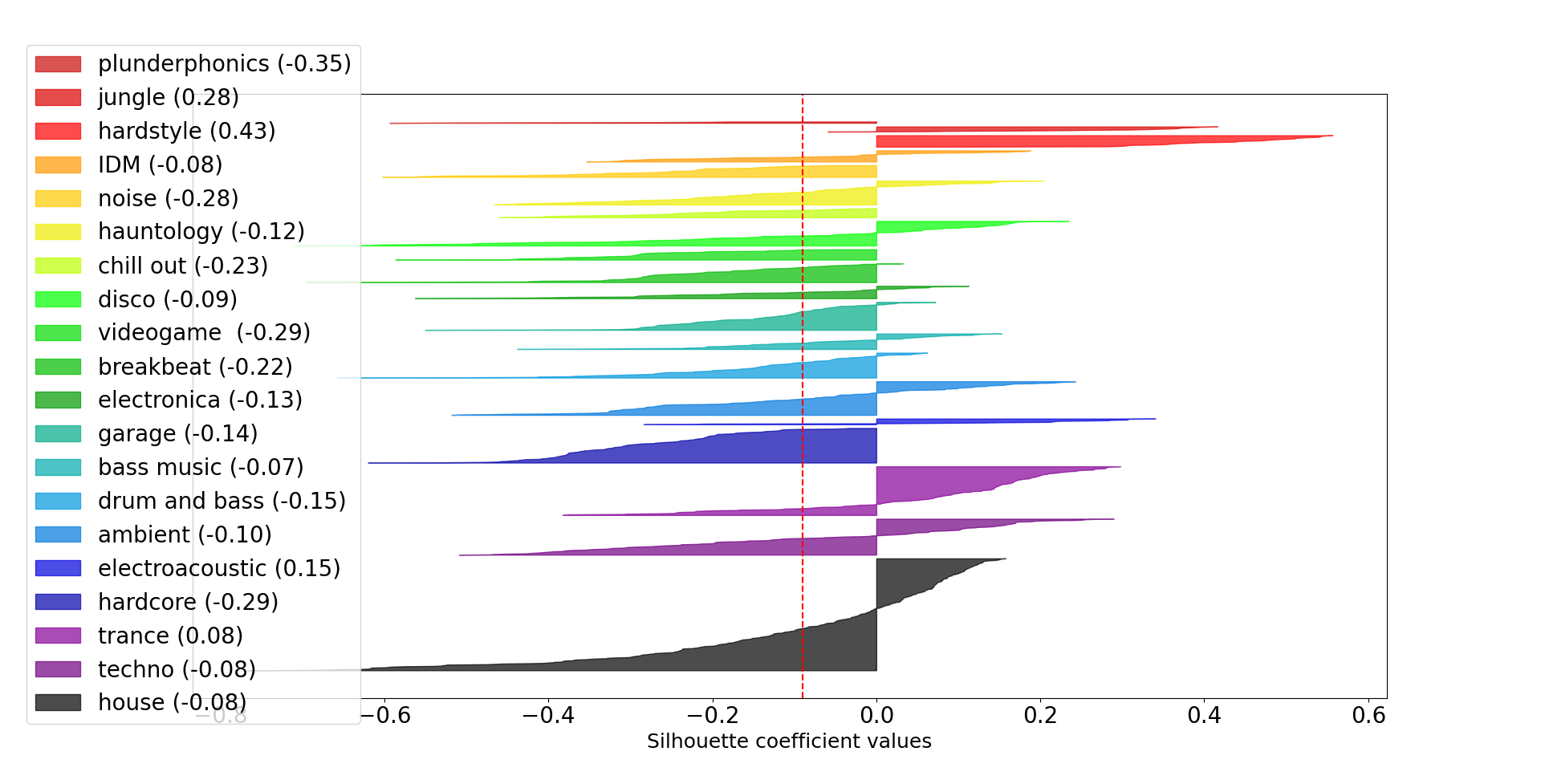}
\caption{Silhouette scores for the tracks clustered by genre. The dashed line indicates the average Silhouette score over all the genres. In the legend, the values in the parenthesis indicate the average genre score.}
\label{fig:5_5}
\end{figure}

In brief, Silhouette analysis measures how much an item of a cluster is similar to other items of its own cluster compared to the ones of other clusters. 
It ranges from -1 to 1, where negative values indicate that an item has been poorly clustered, while positive values mean that an item has been properly matched. 
Given the high dimensionality of the embeddings, we chose to use cosine similarity to measure the distance between items. 
An example of silhouette scores for the 20 genre clusters is reported in Figure \ref{fig:5_5}, computed using the cosine distance between the \textit{EffNet-Discogs} embeddings. 
The score of each genre is computed by averaging the scores of its corresponding tracks. 
Among the others, we see that \textit{Jungle} tracks are well-clustered together (.28), while on the contrary \textit{Hardcore} tracks seem not (--.29). Other genres such as \textit{House} have almost half tracks well-clustered while the other half poorly, obtaining an average score around zero (--.08). 

Silhouette analysis is commonly performed to validate the output of clustering algorithms, however, in our case we employed it as a tool to validate which of the four models provides us with more consistent embeddings according to our ``ground-truth'' labels. 
Averaging the silhouette scores over all the twenty genres, we obtained the best score using the \textit{EffNet-Discogs} model (--.09), followed by \textit{VGGish-AudioSet} (--.11) and then \textit{MusicNN} (\textit{MSD}: --.16, \textit{MTT}: --.17). 
Whilst the difference between models seems not significant, we believe that the use of the Discogs dataset may have boosted the performance of the EfficientNet architecture in creating audio representations which better reflect differences between Electronic Music tracks. 
Indeed, born originally as an EM database, Discogs is a huge source of knowledge of such culture. 
In order to properly compare the different models, each architecture should have been trained with the Discogs database, a task left to practitioners interested in understanding better how deep representations behave with EM tracks. 
From now on, when mentioning embeddings we implicitly refer to the ones generated by the \textbf{EffNet-Discogs} model.

\begin{figure}[t!]
\centering
\includegraphics[width=1\textwidth, ]{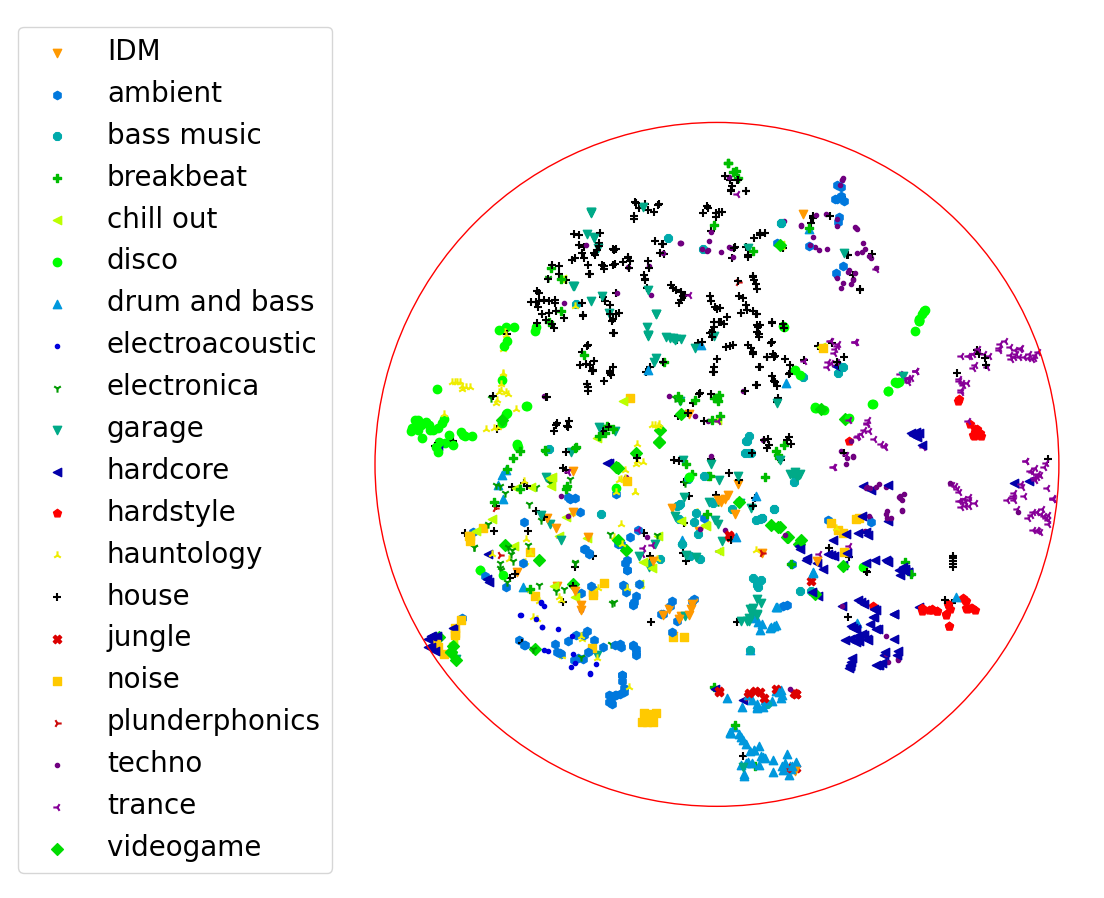}
\caption{Two-dimensional t-SNE projection of candidate tracks' embedding.}
\label{fig:5_6}
\end{figure}

Figure \ref{fig:5_6} displays a two-dimensional representation of the embedded space obtained with this model. 
In the centre of the circle, we notice a quite messy situation with tracks from different genres placed near each other. 
However, going near the border we see a few clusters more defined, for instance the \textit{Trance} tracks on the right or \textit{Drum and Bass} on the bottom. 
We see also how genres which shared music properties are clustered near each other, such as \textit{Drum and Bass}--\textit{Jungle}, or \textit{Hardcore}--\textit{Hardstyle}. 
From the scatter plot, we may also have an intuition about the logic behind the placing of the points in the space according to characteristics such as the BPM or the softness of the sounds.
However, to interpret in depth the nature of the embeddings, we continue our analysis focusing on a series of hand-crafted features as explained in the next section. 

\subsubsection{Hand-crafted Music Features}
Following a first exploration of the embeddings, we scrutinised in depth their relationship with four music features: \textit{tempo}, \textit{danceability}, \textit{acousticness} and \textit{instrumentalness}.
The reason why we selected these features is twofold. 
First, the embeddings seemed to be in some way informative of the distribution of those. 
For instance, the tempo apparently decreased going from bottom to top, while the danceability in the opposite way. 
Second, by using these features we had the opportunity to verify the reliability of Essentia's feature extractor\footnote{\url{https://essentia.upf.edu/streaming_extractor_music.html}} and the Spotify API. 
Both present advantages and disadvantages, and according to the available resources one could prefer one method rather than the other.

Indeed, a main drawback of Essentia is the need for the audio track file to extract the features, while using the Spotify API just with the track ID it is possible to access several audio features. 
However, Spotify algorithms are proprietary, while being Essentia open source it is possible to verify the exact functioning of the extraction process. 
Eventually, we checked the Pearson correlation coefficient ($\rho$)  between the features extracted with Essentia and Spotify, and using the tracks in our dataset we found a positive correlation: 
\textit{tempo} ($\rho=.29$, $p<.01$), \textit{instrumentalness} ($\rho=.42$, $p<.01$), \textit{danceability} ($\rho=.50$, $p<.01$), and \textit{acousticness} ($\rho=.61$, $p<.01$).
Therefore, in terms of analysis no particular difference should emerge if using one method rather than the other. 
From now on, when referring to features we implicitly refer to the ones extracted with \textbf{Essentia}.

\begin{figure}[t!]
\centering
\includegraphics[width=1\textwidth]{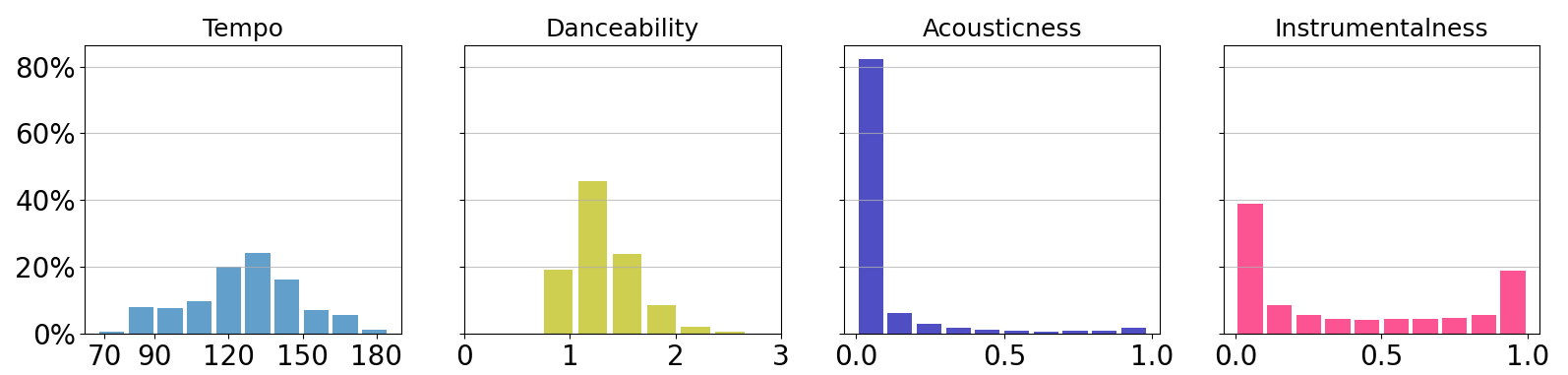}
\caption{Histograms of candidate tracks' feature distributions.}
\label{fig:5_7}
\end{figure}

The track features' distribution (Figure \ref{fig:5_7}) highlight some characteristics of Electronic Music in the dataset. 
First, most of the tracks have a tempo of between 120 and 150 BPM, with some outliers over 160 and under 90 BPM. 
It is worth noting that tempo estimation algorithms suffer from the problem of the so-called \textit{octave errors} i.e. assigning 80 instead of 160 BPM, or vice versa \citep{Schreiber2020}, hence some of these outliers could be a result of these errors. 
In terms of danceability, the extractor returns values between 0 and 3, where higher values mean more danceable. We have the majority of tracks with values between 1 and 2, with 20\% scoring less than 1 and a little percentage scoring more than 2. 
Acousticness and instrumentalness are computed as probabilities ranging from 0 to 1. 
In the case of the former, 0 means almost certainly no presence of acoustic instruments while 1 is the opposite scenario. 
For the latter, 0 means almost certainly the presence of a singing voice, while 1 the absence of singing voice parts. 
In our dataset, it is not surprising to observe that the majority of the tracks are classified as non-acoustic, while we see that the presence of singing voices is less skewed in comparison with the acousticness. 
Computing the correlation between the four features, we found that danceability is negatively correlated with both acousticness and instrumentalness ($\rho= -.44$, $p<.01$ and $\rho= -.66$, $p<.01$), meaning that the more danceable tracks in the dataset are the ones without acoustic instrumentation but with singing parts. 
Instead, these two latter features were positively correlated among them ($\rho=.31$, $p<.01$).

As the last step before defining the diversification strategy, we explored the link between the embeddings and the hand-crafted features. 
In order to do that, we first divided into equally-sized blocks the 2-dimensional projection of the embedded space created (see Figure \ref{fig:5_6}). 
Then, we assigned each track to a block according to its position in the space. Finally, in each block we averaged the feature values of its tracks. 
Because of the non-uniform density of the embedded space, tracks were not equally distributed among blocks. 
\begin{figure}[b!]
\centering
\includegraphics[width=1\textwidth, ]{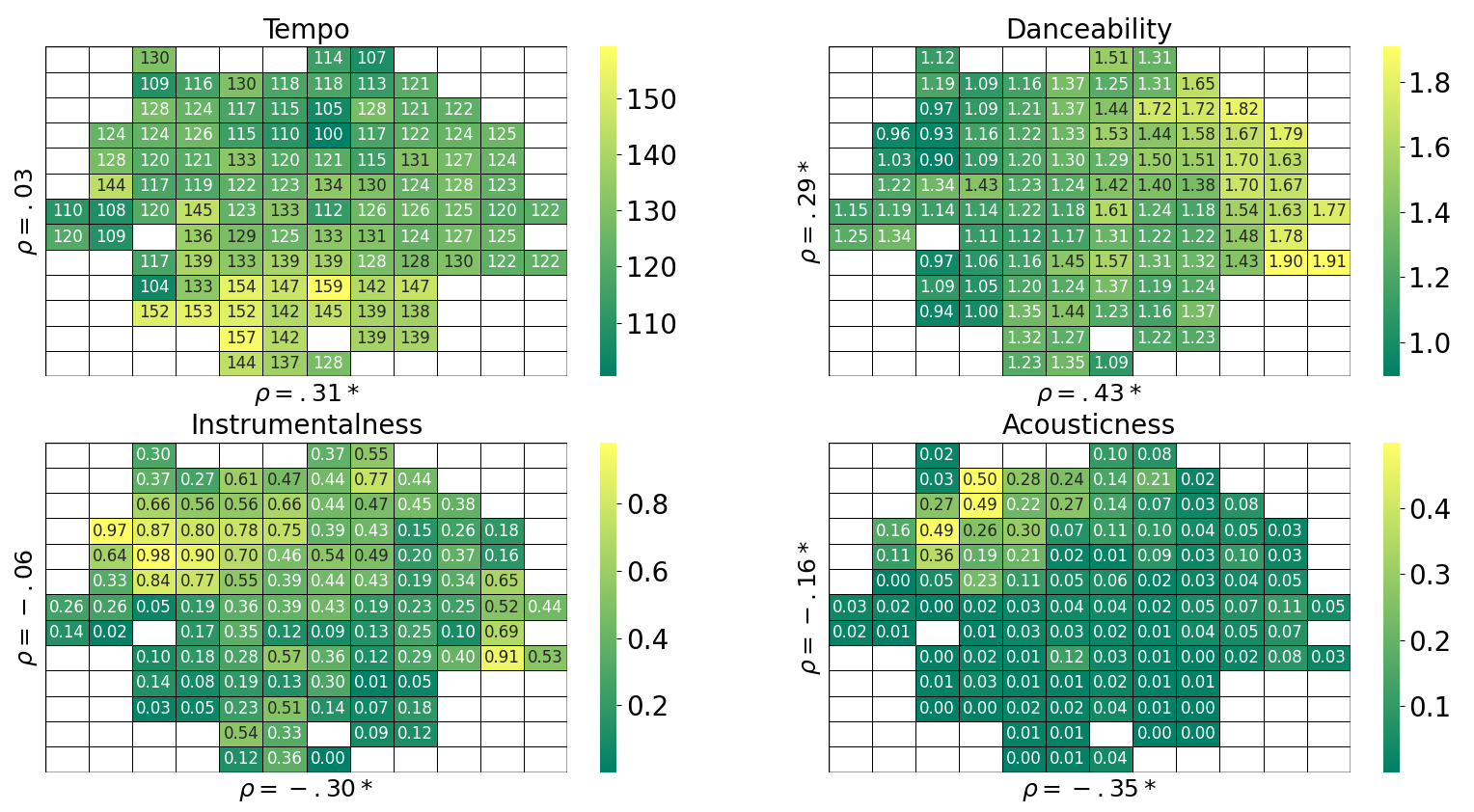}
\caption{Embeddings-features block distribution. The asterisk indicates correlation coefficients with $p<.001$.}
\label{fig:5_8}
\end{figure}
Figure \ref{fig:5_8} displays the four block-heatmaps linking the features to the embedding distribution.

We may observe how the embeddings coherently clustered the tracks with regard to the selected features. 
For instance, tracks with extreme tempos (more than 130 BPM and less than 110 BPM) are distributed in the bottom and bottom-left of the embedded space. 
On the contrary, tracks with high danceability are mostly in the left and top-left parts of the heatmap. Acousticness and instrumentalness instead have higher values in the top-right corner.
Further validation can be obtained by looking at the relationship between the heatmaps and the genre clusters. 
For instance, the \textit{Drum and Bass}, \textit{Hardcore} and \textit{Hardstyle} clusters located at the bottom of the plot correspond to the blocks where the tempo is higher. 
The \textit{Techno}, \textit{Trance} and part of the \textit{House} clusters on the left instead are located where the danceability is higher. 
Instrumentalness goes up in correspondence with \textit{Disco} and \textit{Hauntology} clusters, where also the acousticness is quite high. 

This explorative analysis gives an intuition about how the embeddings may have incorporated musical properties of the tracks, and the correlation coefficients between the x- and y-axis and four features confirm some of our hypotheses. 
A positive correlation of the x-axis means that values increase while going from left to right, while in the y-axis from down to top.

\subsection{Diversity-aware Recommendations}\label{sec:05_diversification}
The diversification process for creating the recommendation to which participants were exposed during the study was based on two main criteria. 
First, we aimed at controlling the \textit{inter-list diversity} to present to one group of participants a varied selection of Electronic Music (EM) throughout the 20 listening sessions, while to the other group only a tiny fraction of EM. 
Second, we limit the \textit{intra-list diversity} for both groups, to avoid creating listening sessions too diverse and fragmented to become potentially annoying. 
Especially having study participants not familiar with EM, to ask them to listen in sequence, e.g. to a glitchy \textit{IDM} track and then a soft \textit{Minimal Techn}o track could have negatively impacted their listening experience, consequently affecting the drop-out rate.

Other design choices of the recommendations were the following. First, each list had to be formed by four tracks to limit the length of the listening session. 
Second, each list had to contain tracks with no more than three different genres. 
Indeed, even if in the previous sections we have shown how the tracks' embedding preserves some musical characteristics of EM genres, it is also true that two tracks could be very near in the embedded space even if labelled differently.

\begin{figure}[h!]
\centering
\includegraphics[width=0.82\textwidth]{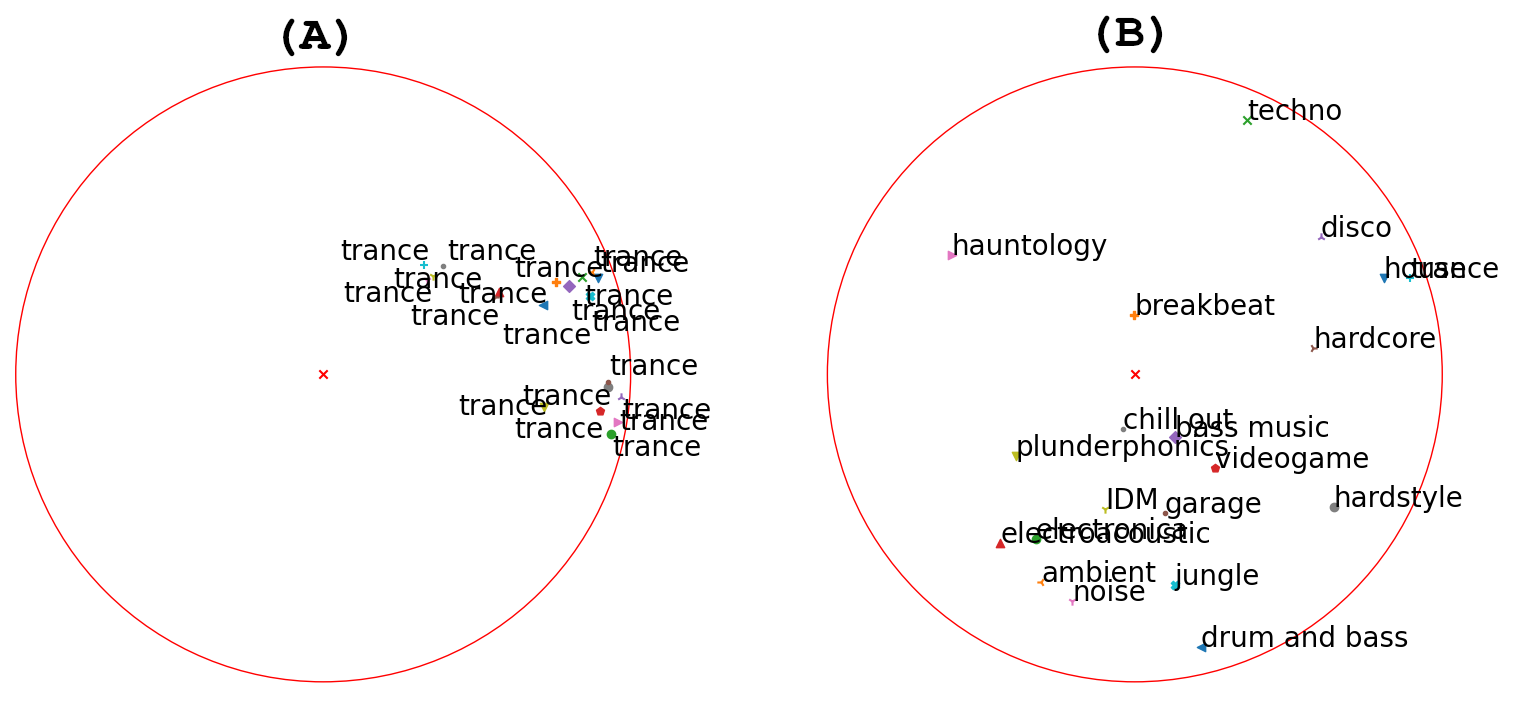}
\caption{Diversification outcomes represented in the 2-dimensional embedded space. Each point represents the average position of a list's tracks. (A) displays the lists with low inter-list diversity (LD), while (B) lists with high inter-list diversity (HD).}
\label{fig:5_9}
\end{figure}

Having in mind these aspects, we implemented the following strategies. 
In order to minimise the intra-list diversity, we found for every track in the dataset the three nearest tracks, according to the pairwise cosine distance between embeddings. 
These quadruples of tracks were the candidate recommendation lists. 
Afterwards, to minimise the inter-list diversity, we selected 20 quadruples from a single genre, while to maximise it we selected one quadruple for each of the 20 genres of the dataset.
Figure \ref{fig:5_9} shows the resulting recommendation lists obtained using these two different diversification processes.

We selected \textit{Trance} as a seed genre of recommendation lists with low inter-list diversity (from now on simply low diversity or LD) for the following reasons. 
First, it has a number of subgenres which ensure some variability between lists, without exceeding like in the case of House (see Table \ref{tab:05_genres}). 
With regard to this aspect, further valid choices could have been \textit{Techno}, \textit{Ambient} or \textit{Hardcore}.
However,  \textit{Trance} tracks have on average a higher silhouette score (Figure \ref{fig:5_5}), meaning that in the embedded space tracks were better clustered than the other genres. 
This aspect was important to ensure to have enough candidates for creating coherent listening sessions with low intra-list diversity. 
Lastly, \textit{Trance} music reflects some stereotypes typical of electronic music: quite danceable, with almost no use of acoustic instruments and a small presence of vocal parts.
Hence, participants exposed to LD recommendations interacted with a stereotypical idea of the EM culture, with few variations in terms of musical properties. 
Even if differences between sessions existed because of different characteristics of the \textit{Trance} subgenres, a certain level of homogeneity was ensured by having selected a single genre to create the recommendations.

\begin{figure}[h!]
\centering
\includegraphics[width=1\textwidth]{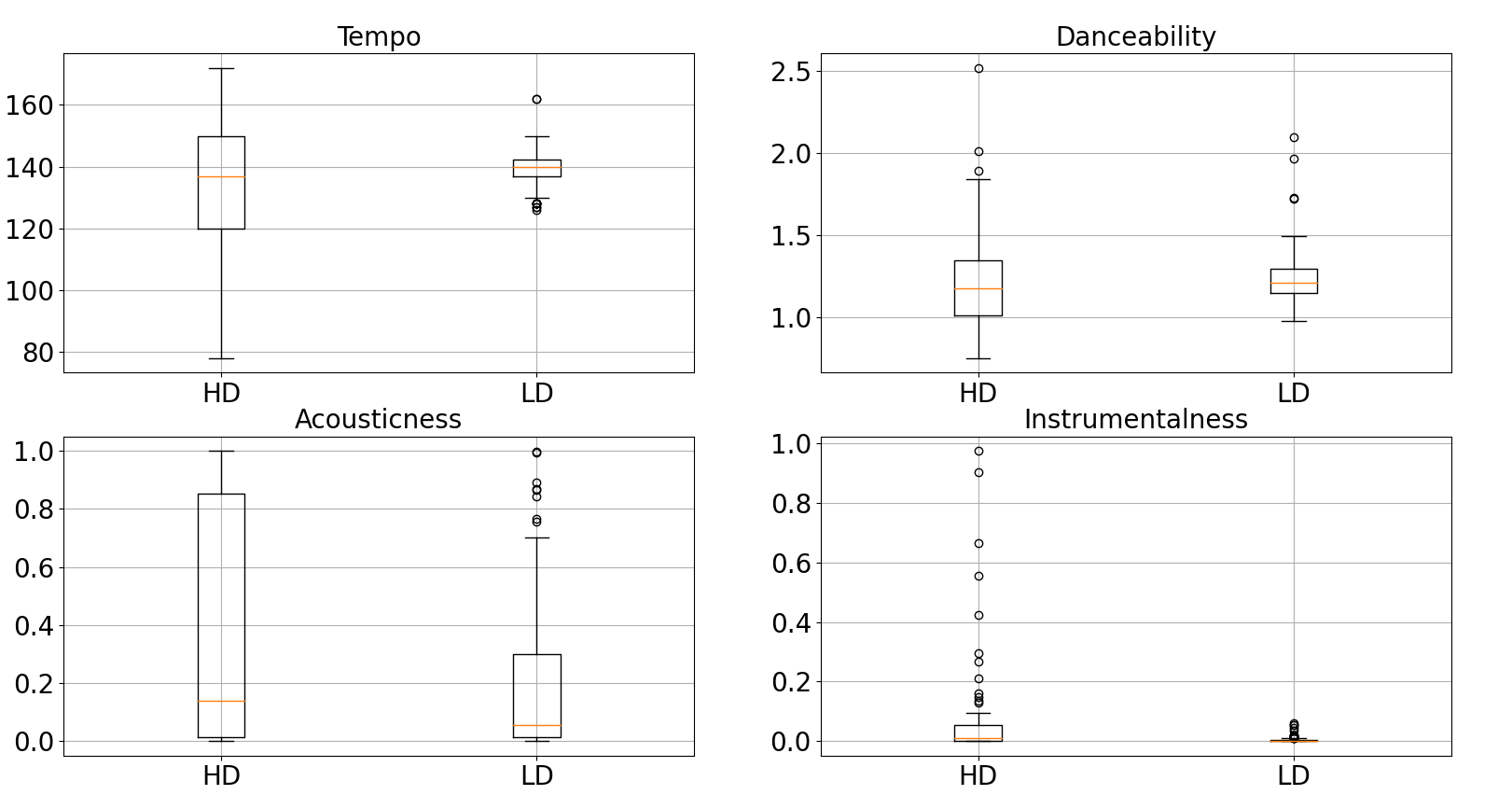}
\caption{Boxplots of the features' distribution for the high diversity (HD) and low diversity (LD) recommendation lists.}
\label{fig:5_10}
\end{figure}

On the contrary, to obtain high inter-list diversity (from now on simply high diversity or HD), we picked a quadruple of tracks for each different genre, which was enough to ensure that participants exposed to such recommendations would explore different facets of EM.
These two diversification strategies effectively led to statistically significant differences between recommendation lists. In detail, we tested these design objectives:
\begin{itemize}
    \item[--] The average \textit{inter-list diversity} should be higher for the HD recommendations in comparison to the LD recommendations.
    \item[--] The average \textit{intra-list diversity} should be equal for the recommendations created using the two diversification strategies.
    \item[--] The average difference between median values of the hand-crafted features should be higher in the HD recommendations than in the LD ones. 
\end{itemize}
According to the statistics presented in Table \ref{tab:design_obj}, the aforementioned design objectives were satisfied. The objectives are further validated by the boxplot presented in Figure \ref{fig:5_10} (Appendix \ref{B_AddTabFig}) which shows the features' distribution.

\begin{table}[h!]
\caption{T-test results comparing high diversity (HD) and low diversity (LD) recommendation lists.}
\centering
\scalebox{1}{
  \begin{tabular}{lrrrrrr}
    \toprule
    &T-stat&df&p-val&CI 95\%&cohen-d&power\\
    \midrule
    Inter-list Div &22.52&378&$<$0.01&[0.27, 0.33]&2.31&1.00\\
    Intra-list Div &2.25&38&0.03&[0.0, 0.07]&0.71&0.59 \\
    Tempo& 9.92&378&$<$0.01&[7.83, 11.7]&1.02&1.00\\
    Danceability&10.56&378&$<$0.01&[0.16, 0.23]&1.08&1.00\\
    Acousticness&9.99&378&$<$0.01&[0.06, 0.09]&1.02&1.00\\
    Instrumentalness&6.05&378&$<$0.01&[0.12, 0.23]&0.62&1.00\\
    \bottomrule
  \end{tabular}}
  \label{tab:design_obj}
\end{table}

Thanks to the aforementioned procedure, we created recommendation lists formed by 4 tracks, 20 with high and 20 with low diversity. 
In every list, tracks were selected to minimise their differences according to the distance between their embeddings. 
The final step was to mix the four tracks in each recommendation list to create a single audio file to be included in a listening session. 
To accomplish that, we implemented the following procedure using \textit{Pysox}, a Python wrapper around Sox \citep{Bittner2016}. 
First, we randomly selected a 45-second excerpt for each track in the list. 
The sample rate of each excerpt was converted to 48000Hz and the volume normalised to -3db. 
Then, we joined the excerpts together including a 1-second fade-in and fade-out, to help the listeners recognize the start and the end of each track in the mix. 
At the end of this process, we obtained a 3-minutes mp3 audio for each recommendation list.

A final manual check of each audio was done to ensure that each listening session was properly mixed. 
W also verified that the content in each audio was appropriate for the purpose of the study. 
Indeed, it is not uncommon that lyrics in EM could contain references to sex, use of drugs, or blasphemy. 
Whilst we do not want to advocate for a moral judgement of the artists' forms of expression, still we agreed that it was necessary to remove some tracks to respect every subjectivity that might eventually participate in the study.

\newpage
\section{Additional Tables and Figures}\label{B_AddTabFig}
\begin{figure}[h!]
\centering
\includegraphics[width=0.9\textwidth]{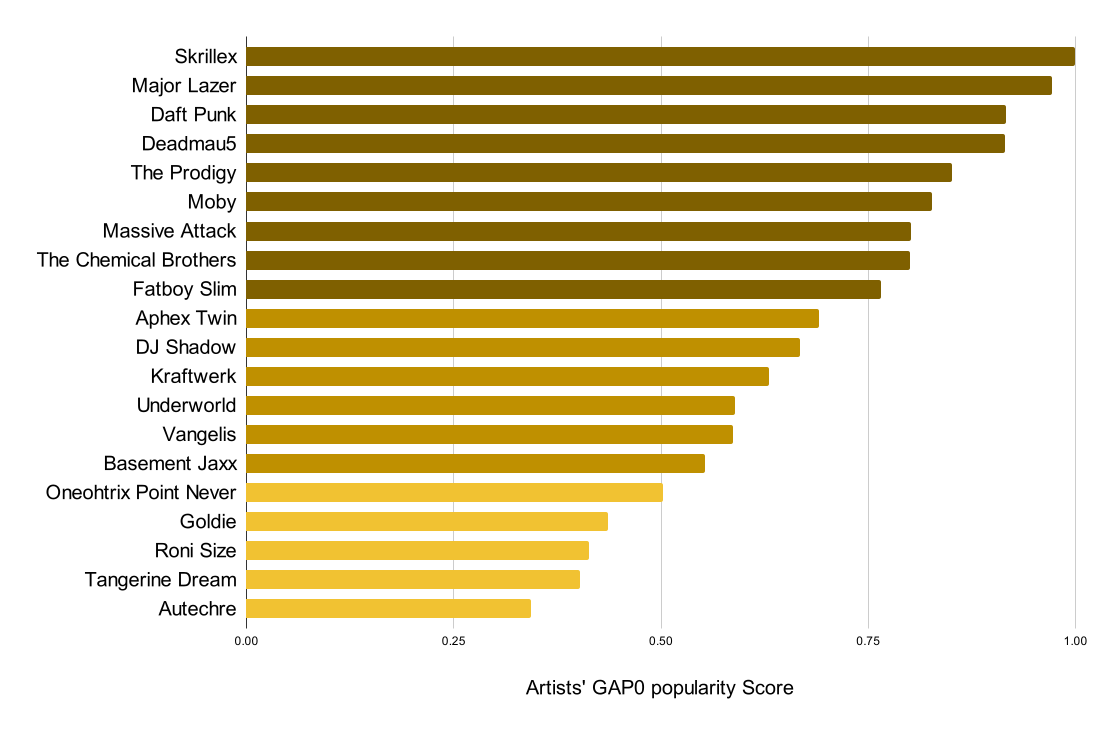}
\caption{List of artists and corresponding GAP0 score for assessing the participants' familiarity with EM.}
\label{fig:5_2}
\end{figure}

\begin{figure}[h!]
\centering
\includegraphics[width=0.9\textwidth]{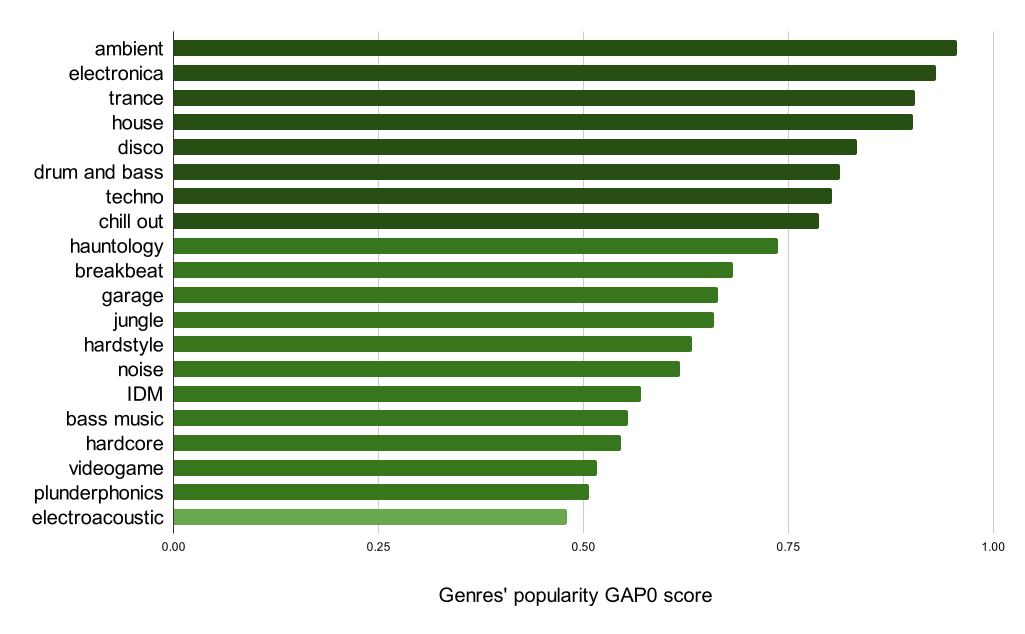}
\caption{List of genres and corresponding GAP0 score for assessing the participants' familiarity with EM.}
\label{fig:5_3}
\end{figure}

\begin{table}[ht!]
\caption{List of genres and subgenres included in the dataset}
\centering
\scalebox{0.95}{
  \begin{tabularx}{\linewidth}{lL}
    \toprule
    \textbf{Genre}&\textbf{Subgenre(s)}\\
    \midrule
    ambient& ambient dub techno, ambient industrial, dark ambient, drone, dungeon synth, illbient, lowercase, new age, new isolationism, space ambient.\\
    bass music& footwork, future bass, kawaii future bass, wave.\\
    breakbeat& big beat, broken beat, hardcore breaks, jersey club, nu skool breaks, progressive breaks.\\
    chill out&downtempo, psydub, trip hop.\\
    disco& boogie, city pop, eurobeat, eurodance, hi-nrg, italo dance, nu disco, post-disco.\\
    drum and bass& atmospheric dnb, darkstep, drumfunk, jump up, liquid funk, neurofunk, sambass.\\
    electroacoustic& acousmatic, electroacoustic composition, electroacoustic improvisation, musique concrete.\\
    electronica& berlin school, folktronica, jazztronica, livetronica.\\
    garage& bassline, brostep, chillstep, dubstep, future garage, grime, speed garage, uk funky, uk garage, wonky.\\
    hardcore&acidcore, breakcore, digital hardcore, doomcore, frenchcore, happy hardcore, industrial hardcore, j-core, makina, speedcore, terrorcore.\\
    hardstyle& euphoric hardstyle, jumpstyle, rawstyle.\\
    hauntology& chillwave, darksynth, future funk, hardvapour, sovietwave, synthwave, vaporwave.\\
    house& acid house, afro house, amapiano, ambient house, bass house, brazilian bass, chicago house, complextro, deep euro house, deep house, disco house, diva house, dutch house, electro swing, fidget house, funky house, future house, garage house, ghettotech, gqom, hard bass, italo house, jazz house, kwaito, latin house, melbourne bounce, microhouse, moombahton, outsider house, progressive house, slap house, soulful house, tech house, tribal house.\\
    IDM& drill and bass, glitch, glitch hop.\\
    jungle&ragga jungle.\\
    noise& death industrial, japanoise, power electronics, power noise.\\
    plunderphonics& ---\\
    techno&acid techno, ambient techno, berlin minimal techno, bleep techno, detroit techno, dub techno, hard techno, industrial techno, minimal techno, raggatek, schranz.\\
    trance& acid trance, dark psytrance, dream trance, full on, goa trance, hands up, hard trance, nitzhonot, progressive psytrance, progressive trance, psychedelic trance, suomisaundi, tech trance, uplifting trance, vocal trance.\\
    videogame& bitpop, chiptune, nintendocore, skweee.\\
    \bottomrule
  \end{tabularx}}
  \label{tab:05_genres}
\end{table}


\begin{figure}[h!]
\centering
\includegraphics[width=1\textwidth]{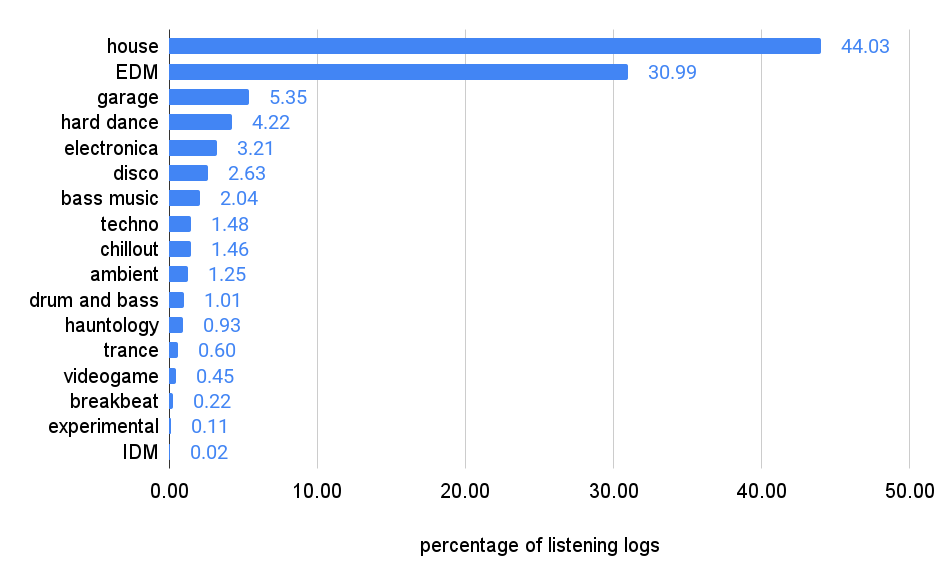}
\caption{Genres ranked by popularity in the study participants' listening logs.}
\label{fig:5_13}
\end{figure}


\begin{figure}[h!]
\centering
\includegraphics[width=1\textwidth]{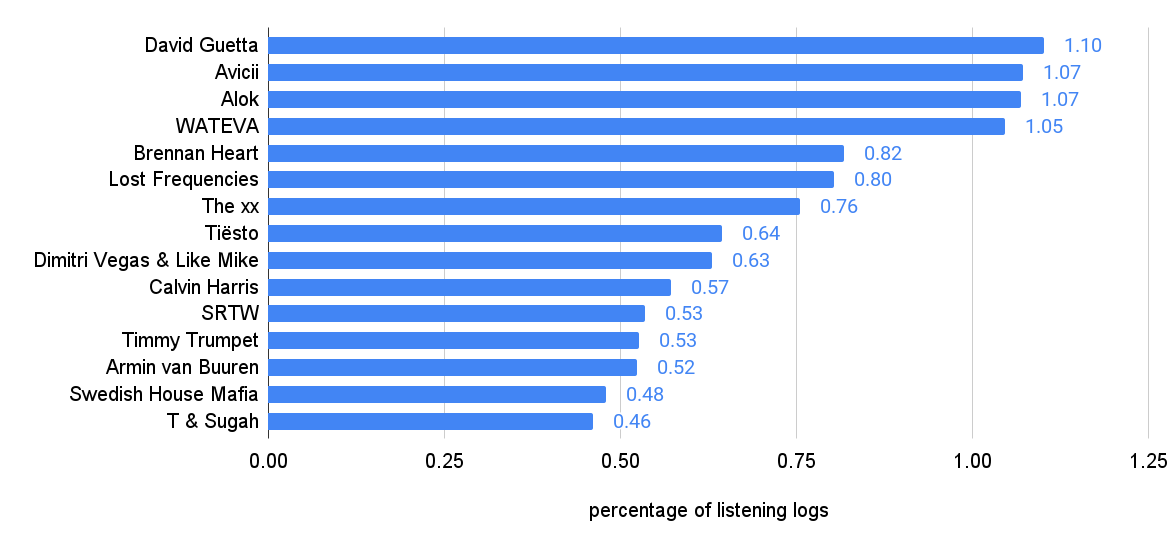}
\caption{Top artists ranked by popularity in the study participants' listening logs.}
\label{fig:5_15}
\end{figure}

\begin{figure}
\centering
\includegraphics[width=\textwidth]{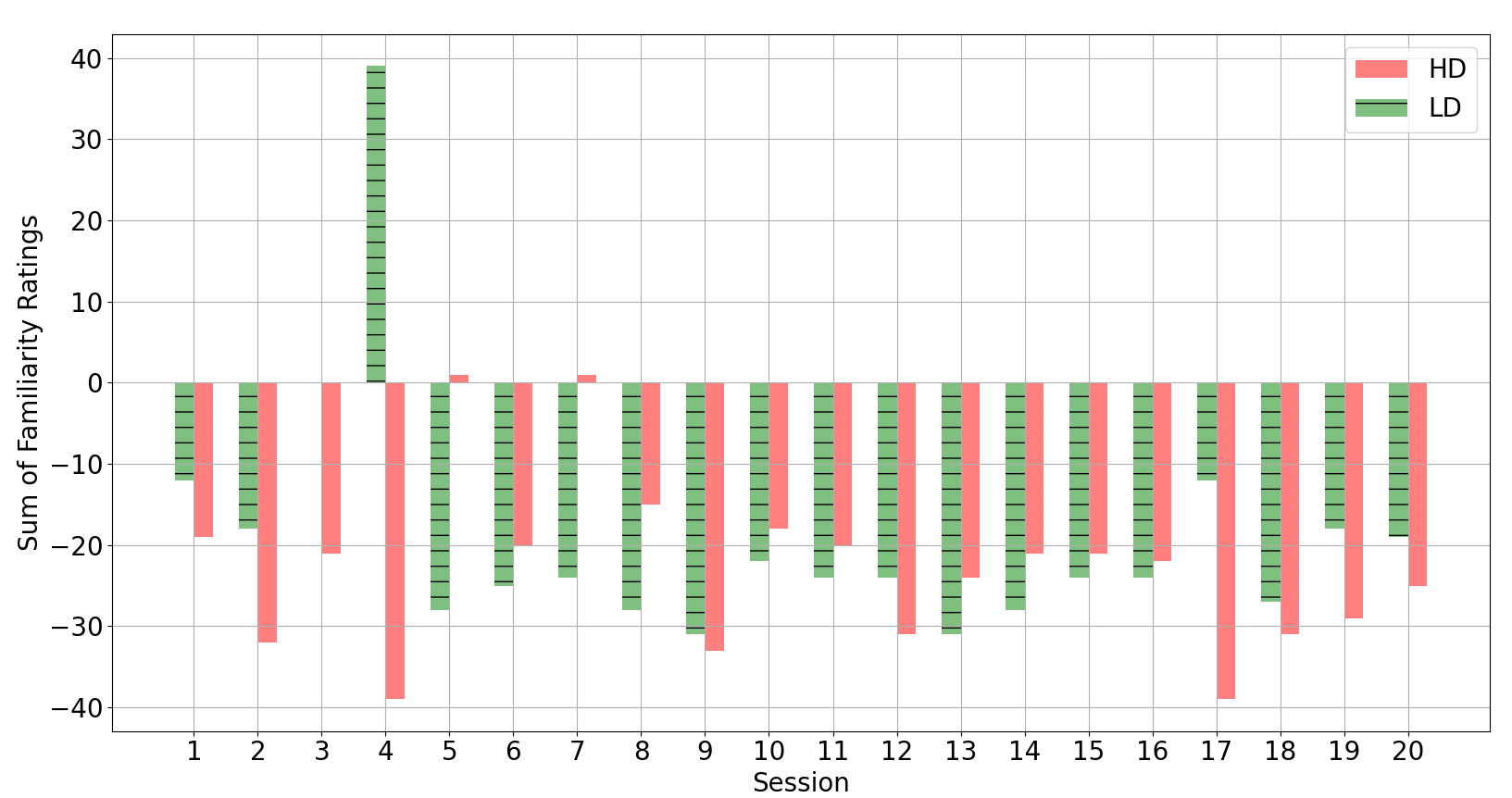}
\caption{Distribution of familiarity ratings.}
\label{fig:5_19}
\end{figure}

\begin{figure}
\centering
\includegraphics[width=\textwidth]{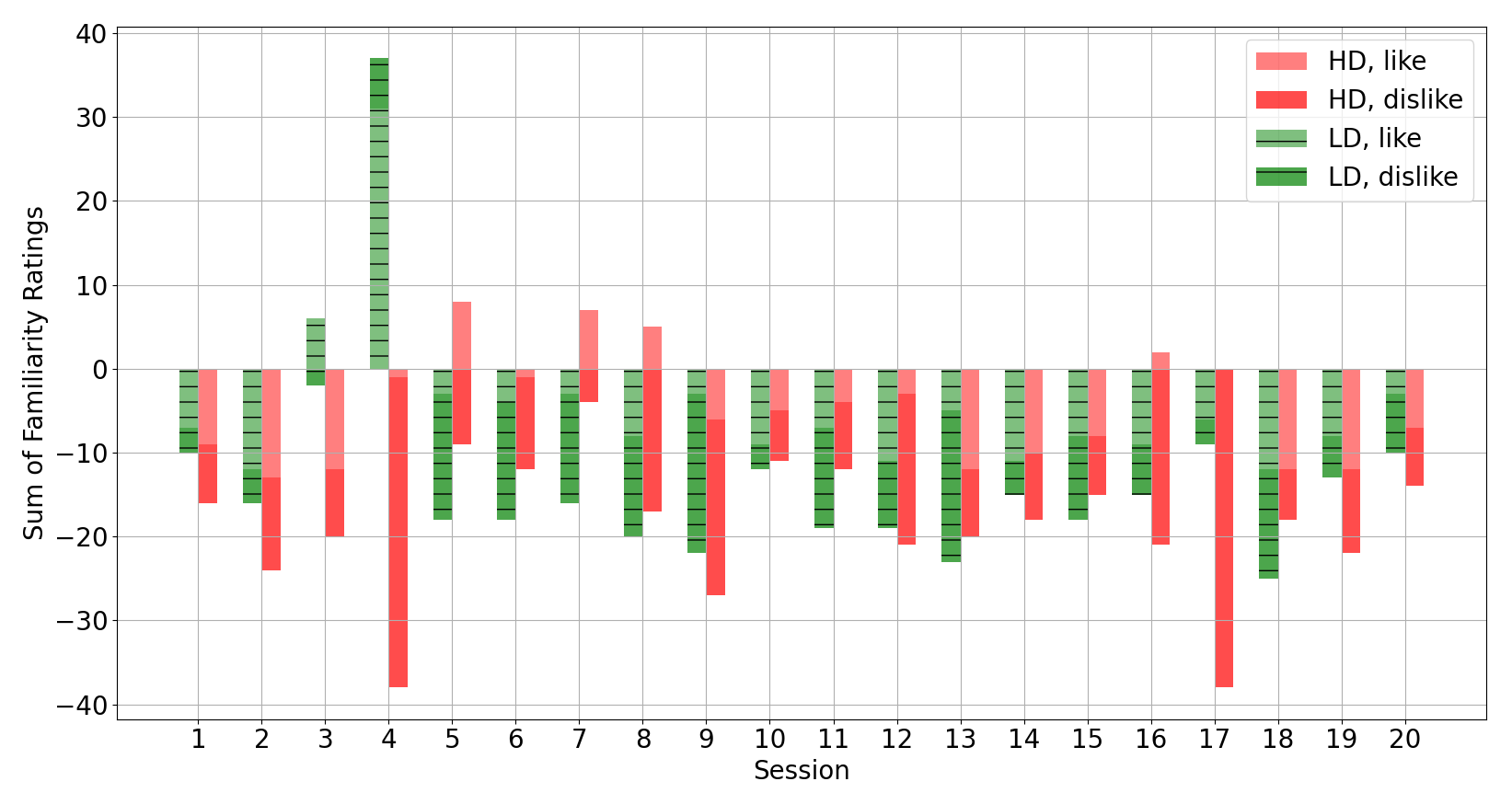}
\caption{Distribution of familiarity ratings split among participants who liked the session (\textit{light bar}) and participants who disliked (\textit{dark bar}).}
\label{fig:5_20}
\end{figure}

\begin{figure}
\centering
\includegraphics[width=\textwidth]{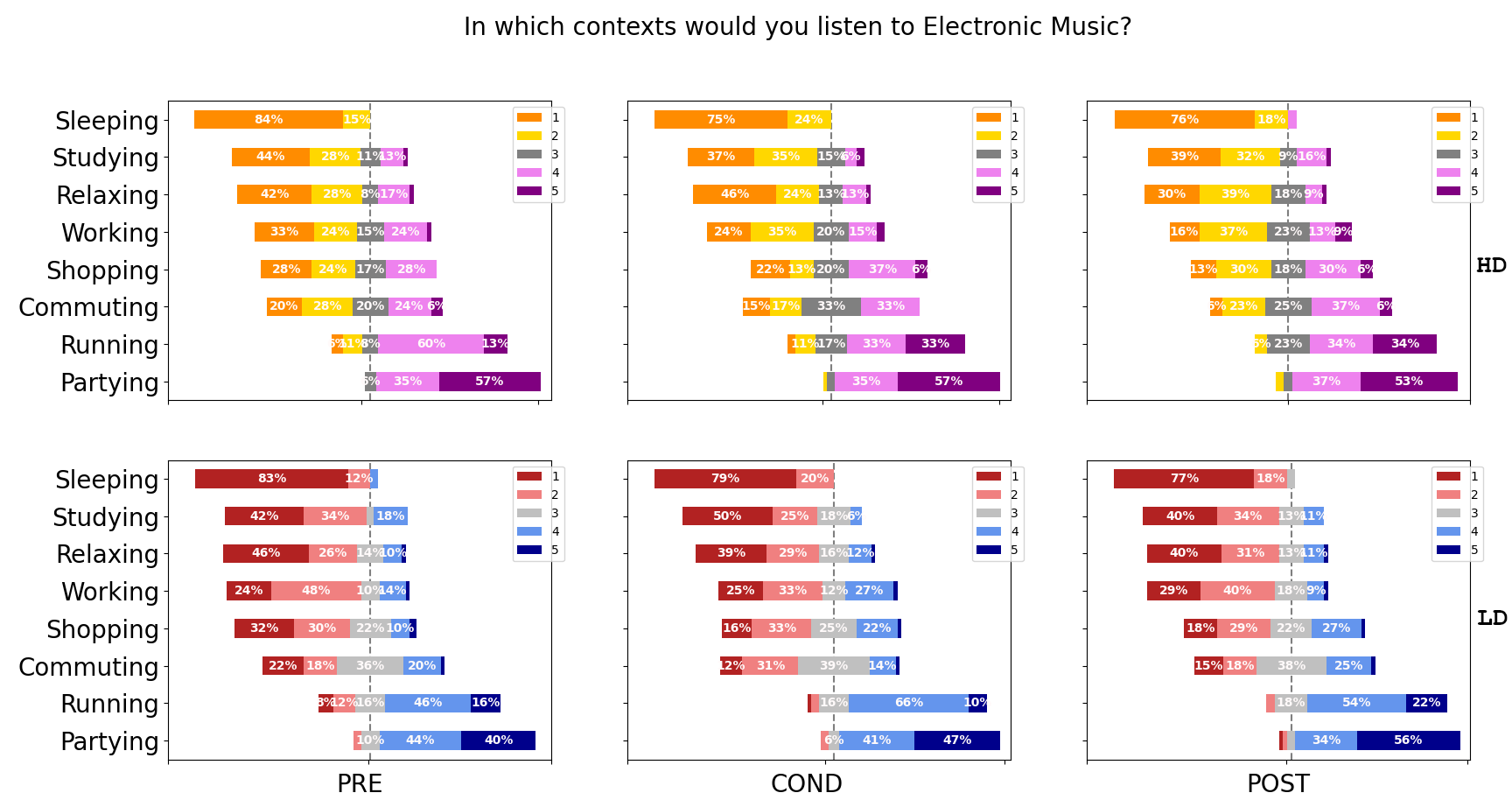}
\caption{Distribution of participants' ratings of the characteristics associated with listening contexts, at the beginning of the experiment (\textit{PRE}), after the exposure (\textit{COND}), and at the end (\textit{right}), separately for the HD group (\textit{top}) and the LD group (\textit{bottom}). In the legend the values of the Likert-item selected are reported (1: Totally Disagree, 5: Totally agree).}
\label{fig:5_24}
\end{figure}

\begin{figure}
\centering
\includegraphics[width=\textwidth]{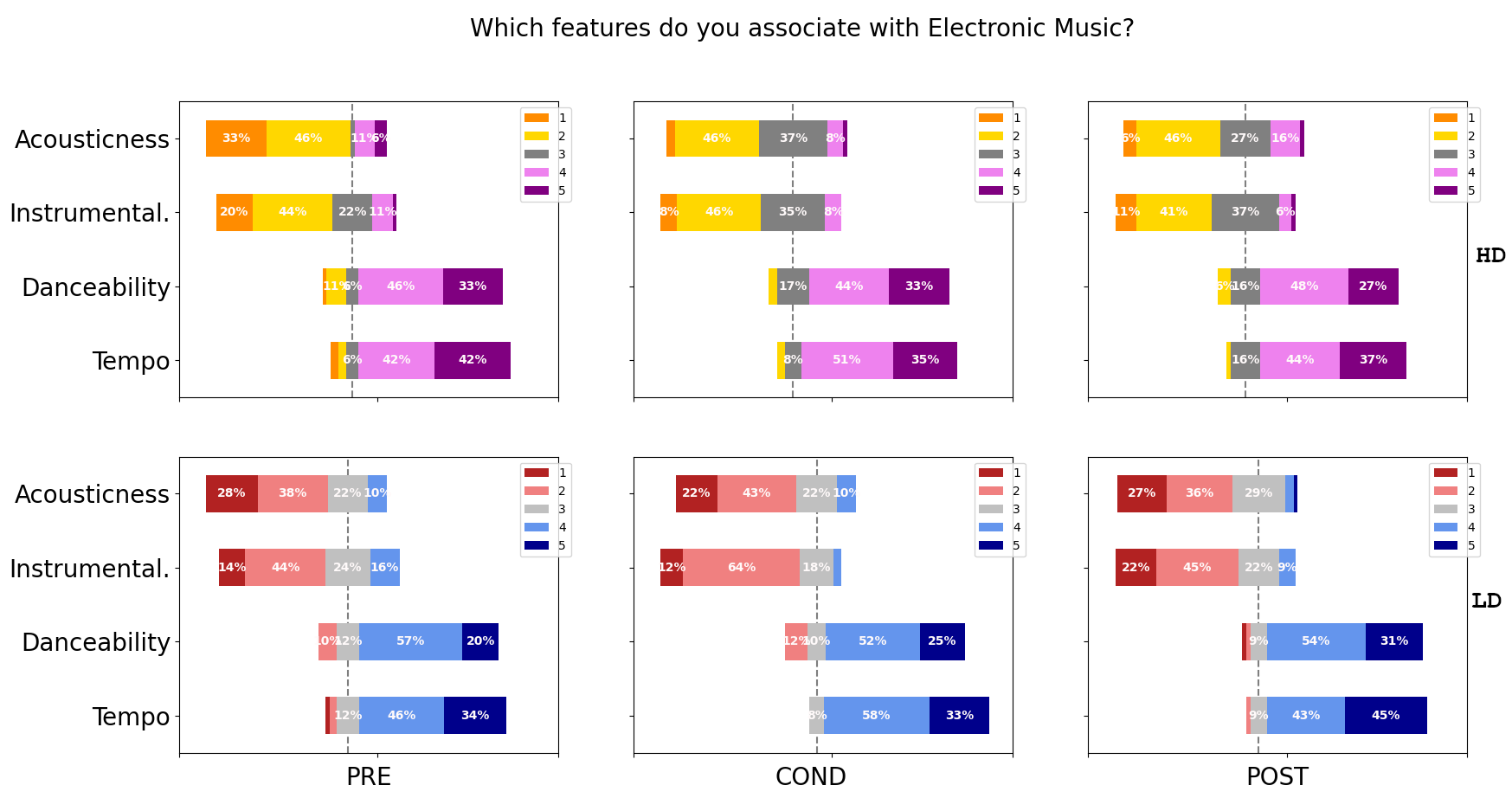}
\caption{Distribution of participants' ratings of the musical characteristics associated with EM tracks, at the beginning of the experiment (\textit{PRE}), after the exposure (\textit{COND}), and at the end (\textit{right}), separately for the HD group (\textit{top}) and the LD group (\textit{bottom)}. The values for the Likert items are: \textit{acousticness} (1: mostly low, 5: mostly high), \textit{instrumentalness} (1: mostly low, 5: mostly high), \textit{danceability} (1: mostly low, 5: mostly high), \textit{tempo} (1: mostly slow, 5: mostly fast).}
\label{fig:5_25}
\end{figure}

\begin{figure}
\centering
\includegraphics[width=\textwidth]{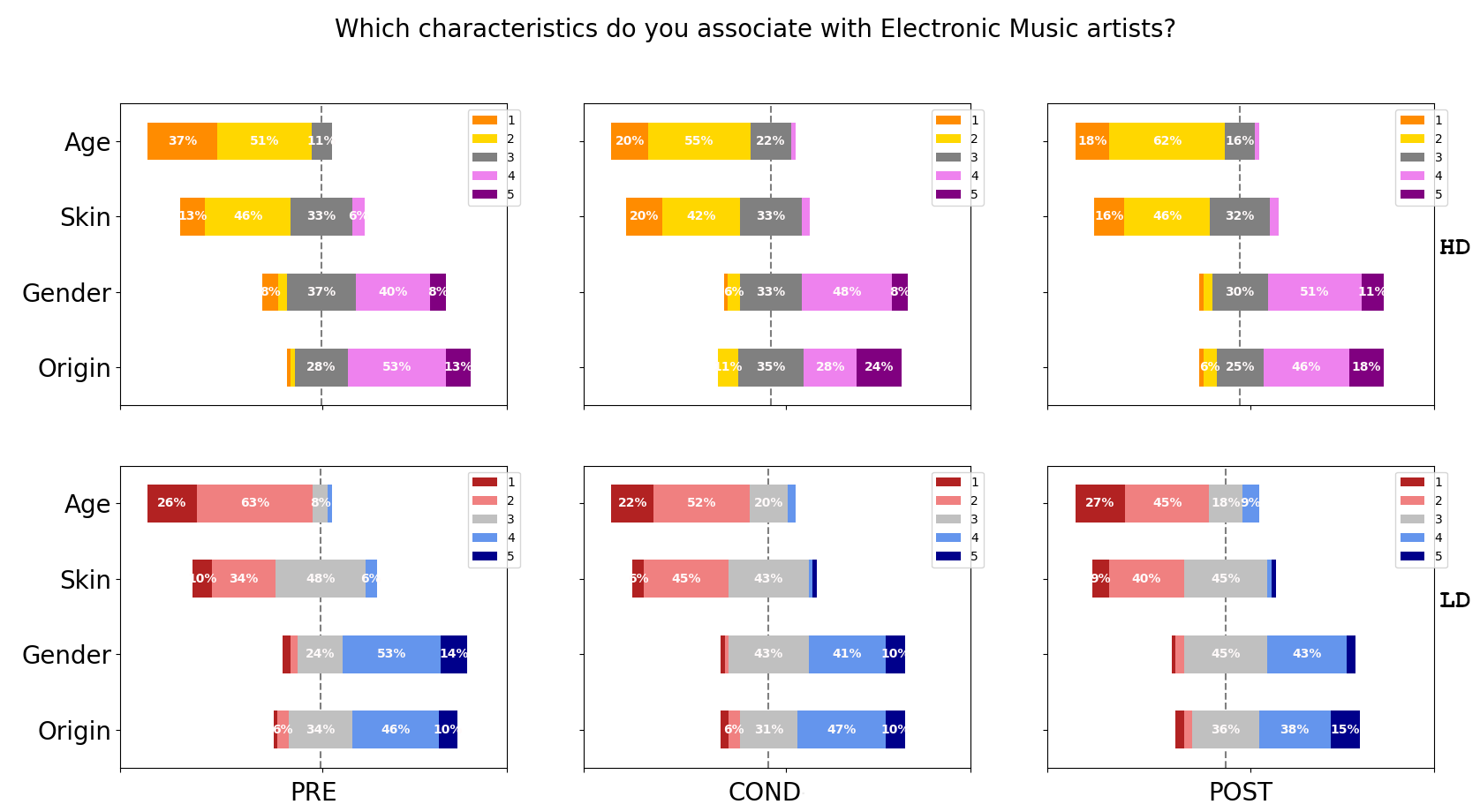}
\caption{Distribution of participants' ratings of the characteristics associated with EM artists, at the beginning of the experiment (\textit{PRE}), after the exposure (\textit{COND}), and at the end (\textit{right}), separately for the HD group (\textit{top}) and the LD group (\textit{bottom}). The values for the Likert items are: \textit{age} (0: mostly under 40, 5: mostly over 40), \textit{skin} (0: mostly white skinned, 5: mostly dark skinned), \textit{gender} (0: mostly women or other gender minorities, 5: mostly men);  \textit{origin} (0: mostly low income / developing countries, 5: mostly high income / developed countries).}
\label{fig:5_26}
\end{figure}

\end{document}